\documentclass[sigconf,nonacm]{acmart}
\AtBeginDocument{%
  }
    
\usepackage{algorithm}
\usepackage{algorithmic}
\usepackage{amsmath}
\usepackage{tcolorbox}
\usepackage{graphicx}
\usepackage{lipsum}
\usepackage{longtable}
\usepackage{soul}
\usepackage{booktabs}
\usepackage{array}
\usepackage{float}
\usepackage{multirow}
\usepackage{colortbl}
\usepackage{subcaption}
\usepackage{graphicx}
\usepackage{subcaption}
\usepackage{enumitem}
\usepackage{svg}

\begin{document}

\title[Wisdom of the LLM Crowd]{Wisdom of the LLM Crowd: A Large Scale Benchmark of Multi-Label U.S. Election-Related Harmful Social Media Content}

\author{Qile Wang}
\orcid{0000-0003-0308-6033}
\affiliation{%
  \institution{University of Delaware}
  \city{Newark}
  \state{DE}
  \country{USA}
}
\email{kylewang@udel.edu}

\author{Prerana Khatiwada}
\affiliation{%
  \institution{University of Delaware}
  \city{Newark}
  \state{DE}
  \country{USA}
}
\orcid{0009-0008-6965-9504}
\email{preranak@udel.edu}

\author{Carolina Coimbra Vieira}
\affiliation{%
  \institution{Max Planck Institute for Demographic Research}
  \city{Rostock}
  \country{Germany}
}
\orcid{0000-0003-3156-4151}
\email{coimbravieira@demogr.mpg.de}

\author{Benjamin E. Bagozzi}
\affiliation{%
  \institution{University of Delaware}
  \city{Newark}
  \state{DE}
  \country{USA}
}
\orcid{0000-0002-6233-6453}
\email{bagozzib@udel.edu}

\author{Kenneth E. Barner}
\affiliation{%
  \institution{University of Delaware}
  \city{Newark}
  \state{DE}
  \country{USA}
}
\orcid{0000-0002-0936-7840}
\email{barner@udel.edu}

\author{Matthew Louis Mauriello}
\affiliation{%
  \institution{University of Delaware}
  \city{Newark}
  \state{DE}
  \country{USA}
}
\orcid{0000-0001-5359-6520}
\email{mlm@udel.edu}

\renewcommand{\shortauthors}{Wang \emph{et al.}}

\begin{abstract}
The spread of election misinformation and harmful political content conveys misleading narratives and poses a serious threat to democratic integrity. Detecting harmful content at early stages is essential for understanding and potentially mitigating its downstream spread. In this study, we introduce USE24-XD, a large-scale dataset of nearly 100k posts collected from X (formerly Twitter) during the 2024 U.S. presidential election cycle, enriched with spatio-temporal metadata. 
To substantially reduce the cost of manual annotation while enabling scalable categorization, we employ six large language models (LLMs) to systematically annotate posts across five nuanced categories: Conspiracy, Sensationalism, Hate Speech, Speculation, and Satire. We validate LLM annotations with crowdsourcing (n = 34) and benchmark them against human annotators. Inter-rater reliability analyses show comparable agreement patterns between LLMs and humans, with LLMs exhibiting higher internal consistency and achieving up to 0.90 recall on Speculation. 
We apply a wisdom-of-the-crowd approach across LLMs to aggregate annotations and curate a robust multi-label dataset. 60\% of posts receive at least one label.
We further analyze how human annotator demographics, including political ideology and affiliation, shape labeling behavior, highlighting systematic sources of subjectivity in judgments of harmful content. 
The USE24-XD dataset is publicly released to support future research.
\begin{itemize}
  \item Dataset $\to$ \url{https://github.com/Sensify-Lab/USE24-XD}
\end{itemize}
\end{abstract}

\maketitle

\section{Introduction}
Online harmful content, including misinformation, increasingly undermines modern elections by shaping public opinion, influencing voter behavior, and destabilizing democratic processes \cite{au2022role, al2024harmful, giachanou2020battle, harm}.
This challenge is amplified by the growing volume of online content and the increasing reliance on social media platforms for political information, whose networked diffusion mechanisms facilitate the rapid spread of misleading narratives \cite{moravec2019fake}.
As online harmful content continues to influence elections, a limited number of standardized datasets (\emph{e.g.,} \citep{torabi2019big, kennedy2022repeat, chen2022election2020, raza2024fakewatch, pinto2025tracking}) have been developed to evaluate election-related harmful content.
Many of these efforts are framed using terms such as misinformation and fake news, reflecting prevailing terminology \cite{zeng2023misinformation}, while focusing on specific classes of election-related harms.
These developments underscore the pressing need for robust, election-specific datasets to support the development of more effective systems for detecting harmful content. Raw social media data are often messy and unstructured \cite{velkova2023unstructured, elsayed2019proposed}, underscoring the importance of curated, cleaned datasets for advancing research on online harmful content. However, developing such datasets poses significant challenges. Annotators must invest substantial time and effort to categorize content manually, a process that is costly and prone to inconsistency, especially when classifying ambiguous or subjective material \cite{said2017cost}. Although some researchers have proposed automated annotation using LLMs \cite{tan2024large, su2022selective}, few have systematically compared the performance or evaluated the accuracy of different LLMs when labeling diverse categories of harmful content.

Our goals are twofold: (i) to evaluate the effectiveness of LLMs in classifying potentially harmful or sensitive online content and (ii) to identify patterns, if any, in the use of generative AI as a supportive annotation tool. Further, our study aims to answer the following research questions:\\
\textbf{RQ1:} \textit{How do different LLMs perform in annotating online harmful content, and which model characteristics explain their annotation behavior?}\\
\textbf{RQ2:} \textit{How do human annotators perform in annotating online harmful content, and to what extent do human annotators' demographics and characteristics, such as political ideology, shape the annotation of online harmful content?}\\
\textbf{RQ3:} \textit{To what extent do LLM predictions agree, and how does this agreement compare with human annotators?}\\
To address these questions, we evaluate six leading LLMs with different architectures on social media posts from X\footnote{\url{https://x.com/}} that may contain potentially harmful information across five categories: Conspiracy, Hate Speech, Satire, Sensationalism, and Speculation. 

\begin{figure*}[!htb]
    \centering
    \includegraphics[width=1.5\columnwidth]{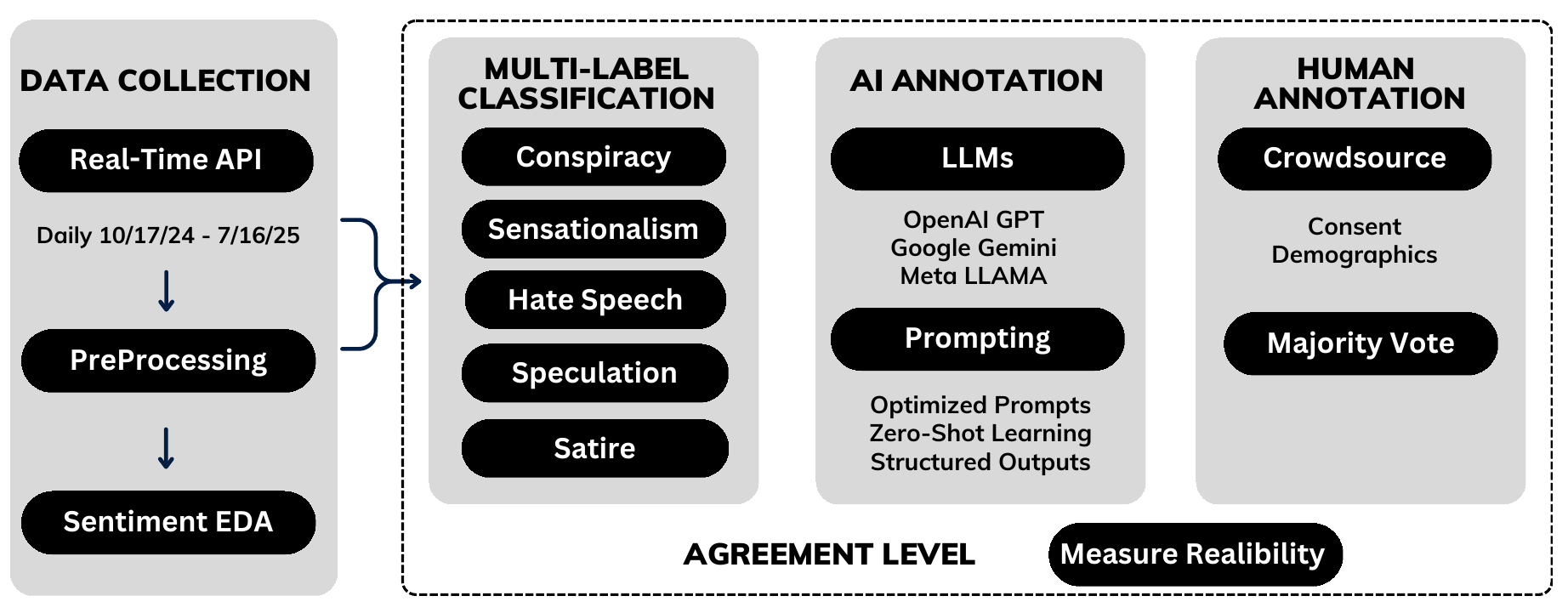}
    \caption{Overall Method for Multi-label Harmful Online Content Annotation}
    \label{fig:overall}
\end{figure*}

As shown in Figure \ref{fig:overall}, we validate a subset via human annotation on a crowdsourcing platform and adopt a wisdom of the crowd approach across the three best-performing LLMs, using majority voting to establish consensus labels. Our approach draws inspiration from annotation-efficient frameworks, such as the vote-k strategy, which uses selective annotation and prompt retrieval to enhance language task performance~\cite{su2022selective}. We observed that LLMs achieved higher inter-rater reliability (mean Krippendorff’s $\alpha = 0.70$ for optimized model combinations) compared to human annotators (mean $\alpha = 0.62$ for the best annotation sets per category). When using human annotations as ground truth, LLM-assisted annotation achieved Recall scores ranging from 0.68 (Satire) to 0.90 (Speculation), demonstrating its potential as a supportive and scalable tool for multi-label annotation of harmful online content.
We further explore the influence of annotators’ demographics on labeling patterns to reveal how social perspectives shape judgments of harmful content.
The main contributions of this paper are summarized as:

$\triangleright$ We identify five categories of harmful online content and present a crowdsourcing study to manually annotate harmful content with multi-labels, analyzing how annotators’ labeling decisions relate to their demographic backgrounds.

$\triangleright$ We introduce a comprehensive evaluation of six state-of-the-art LLMs for the same multi-label harmful content annotation, benchmarking their performance against human annotators.

$\triangleright$ We release the U.S. Election 2024 X.com Dataset (USE24-XD), a recent, well-curated resource designed to advance research on social media and the web.

\section{Related Work}
We review prior work on: (1) election-related online harmful content, often studied through misinformation narratives; and (2) annotation gaps that motivate our scalable LLM-based approach.

\subsection{Online Harmful Content Detection}
Studies on misinformation detection have explored a range of Natural Language Processing (NLP) techniques and reported high performance across many datasets \cite{pelrine2021surprising}. However, efforts to detect election-related misinformation remain in their early stages, particularly on dynamic social media platforms where content is unstructured, fast-moving, and multimodal. Prior work has noted that election-related harms online often extend beyond mere factual inaccuracies, intersecting with broader forms of harmful content such as hate speech, harassment, and other socially consequential expressions \cite{wardle2017information, tercova2025digital}.
Collecting data from platforms like X is especially challenging due to API limitations, data volume, and the need for careful filtering, which explains why relatively few studies focus on this domain today. Existing datasets and models for misinformation detection often rely on heuristic-based or automated labeling schemes, such as keyword matching, hashtags, or user metadata \cite{chen2022election2020}. While convenient, these approaches can oversimplify nuanced content and produce noisy labels. For example, some Reddit-based datasets infer misinformation from subreddit themes or selected comments \cite{nakamura2020fakeddit}, which may not capture the broader discourse. Similarly, weakly supervised approaches on TikTok, such as the analysis of 1.8 million videos during the 2024 U.S. Presidential Election \cite{pinto2024tracking}, reveal political clusters and misinformation trends but lack fine-grained, election-specific labels and do not account for other forms of harm embedded in political content.

Building on prior work largely focused on misinformation detection, recent studies increasingly emphasize the need for scalable and nuanced approaches to address the complexity of election related content on modern social media platforms \cite{aimeur2023fake}.
Prior work, such as FakeWatch
\cite{raza2024fakewatch} focuses on identifying misinformation in North American election-related news articles using both traditional machine learning models and state-of-the-art language models. While effective, these studies highlight the trade-off between computational efficiency and model complexity, underscoring the continued relevance of classical ML approaches alongside deep learning. Although several datasets \cite{10.1145/3485447.3512126} and models exist for misinformation detection, few offer rigorously validated annotations across multiple misinformation sub-topics at scale. To address these limitations, our approach builds on the misinformation literature and leverages LLMs with consensus-based techniques to produce scalable, consistent annotations of online harmful content. Compared to traditional manual or heuristic methods, this strategy captures nuanced discourse more effectively. It enables systematic analysis of election-related harms, encompassing misinformation and related social and interpretive harms (\emph{e.g.,} hate speech and satire).

\subsection{AI \& Human Annotation}
LLMs have become increasingly popular for various applications, such as stance detection \cite{10.1145/3701716.3715521}, review classification \cite{wang2025leveraging}, and thematic analysis \cite{wang2025lata}.
Recent work has shown that LLMs, such as ChatGPT and Claude, can approach human-level performance in rationale annotations \cite{nguyen2024human} or multi-label text classification, with ChatGPT-4o emerging as the most reliable model when classifying Spanish-language newspaper articles on immigration \cite{tripp2025benchmarking}.
ChatGPT also outperforms both crowd workers and trained annotators in text annotation tasks such as relevance, stance, topic, and frame detection, achieving higher accuracy and greater intercoder agreement while reducing annotation costs by thirty times compared to Amazon Mechanical Turk \cite{gilardi2023chatgpt}.
Beyond general text annotation, LLMs have also been applied to specialized domains. \citet{hamilton2024gpt} scaled propaganda-feature annotation with a GPT-3.5-assisted pipeline, achieving near-GPT-4 accuracy at lower cost. 
Similarly, \citet{choi2024automated} developed FACT-GPT, an LLM-driven claim-matching framework to support fact-checkers in detecting recycled misinformation, demonstrating that fine-tuned LLMs can rival larger models and align closely with human annotations.

Despite these promising results, LLM outputs can still vary in quality. To address this, a growing line of research has explored the strategic integration of LLMs with human annotation, leveraging the LLM’s verbalized confidence to request human annotators for low-confidence cases selectively \cite{gligoric2025can, wang2024human}. 
Alternatively, \citet{wang2024explainable} proposed a framework for fake news detection that generates evidence from opposing crowd opinions, uses LLMs to produce justifications for each side, and determines veracity through a defense-based reasoning process. In this work, we move beyond relying on a single LLM and leverage \textit{the Wisdom of Crowds} \cite{welinder2010multidimensional} to generate misinformation annotations across multiple LLMs, thereby improving annotation quality.
Inspired by these developments, we treat each LLM in our study as an independent generative AI annotator and aggregate its predictions using a majority vote strategy. To evaluate LLM performance in depth, we employed random sampling and used crowdsourcing for human verification.

\section{Methodology}
Our methodology (Figure \ref{fig:overall}) consists of three main steps: data collection, multi-label annotation of potentially harmful online content using LLMs, and human annotation using crowdsourcing.

\subsection{Dataset Collection (USE24-XD)}
We crafted the \textbf{U.S. Election 2024 X.com Dataset (USE24-XD)\footnote{\url{https://github.com/Sensify-Lab/USE24-XD}}} by collecting nearly 100k posts from X.com between October 17, 2024, and July 16, 2025, covering the U.S. presidential election period.
With the academic API discontinued and full-archive access restricted to enterprise subscribers, we relied on a basic-tier subscription that only allowed real-time collection of up to 10,000 posts per month, later increased to 15,000 posts per month after the API update in November 2024. To optimize the collection rate, we set daily targets to distribute API usage evenly, collecting roughly four times per day to capture posts throughout the day.
Our search operators identify posts containing both \textit{``election''} and \textit{``2024''} to capture broad election-related discourse without topical restrictions.
The query is also filtered to include only original posts. The dataset includes posts from users worldwide and is not limited to any specific geographic region, as place-tagged posts are under 1\% and geo coordinates-tagged posts make up less than 0.1\% of posts on X. \cite{kruspe2021changes}.

To approximate user location, we included the “user location” field provided in the post metadata. However, this location information is user-reported and, therefore, serves only as an estimate rather than a verified geographic indicator. 
Our dataset includes additional metadata provided by the API, such as public engagement metrics and content sensitivity. We removed duplicates and posts with fewer than five words, as they are challenging to label and often consist only of links or references to other media. We then performed post-processing (\emph{e.g.,} removing URLs, mentions, hashtags, and punctuation, and converting text to lowercase) and appended additional columns, including \texttt{word\_count}, \texttt{clean\_text}, sentiment analysis, and topic features. These additional features are developed exclusively for exploratory data analysis, and the results are presented in Appendix \ref{EDA}. Detailed dataset statistics are presented in Table~\ref{datastat}.

\begin{table}[!htbp]
\footnotesize
\centering
\caption{Dataset USE24-XD Characteristics}
\begin{tabular}{rm{4.85cm}}
\toprule
\textbf{Metric} & \textbf{USE24-XD} \\ 
\midrule
        Post Creation Date & October 17, 2024 - July 16, 2025\\
        Total Posts after cleaning (n) & 97,696 \\ 
        Sensitive Post (n) & 1096 (1.12\%) \\ 
        Verified Post (n) & 4623 (4.73\%) \\ 
        Unique Users (n) & 58824 \\
\midrule
        Repost Count (Avg $\pm$ SD) & 2.83 $\pm$ 58.18 \\ 
        Like Count (Avg $\pm$ SD)  & 11.79 $\pm$ 226.03 \\ 
        Impression (Avg $\pm$ SD)  & 517.15 $\pm$ 6870.41 \\ 
        Word Count (Avg $\pm$ SD)  & 31.79 $\pm$ 13.81 \\
        Top 5 Verified Account & \scriptsize{grok, WashTimes, Reuters, NEWSMAX, AskPerplexity}\\
\bottomrule
\end{tabular}
\label{datastat}
\end{table}

\subsection{Annotation Categories}

Based on theoretical and empirical research in political discourse and harmful online content, we defined five categories for multi-label binary classification.
Focusing on the text modality from the dataset, we aimed to capture distinct but overlapping modes of manipulation and distortion. Two researchers conducted a literature review of prior research related to information quality and online content to consolidate related labels into broader categories. For example, we grouped ``Hoaxes” and ``False Political Claims” under ``Speculative or Unverified Claims,” and merged “Inflammatory Rhetoric” with ``Derogatory Speech and Defamation” to form ``Hate Speech”.
The final annotation framework comprises five independent binary classification tasks, each described below by its definition and the information it captures. A post may not belong to any category, or it may belong to multiple categories. These labels fall under the broader category of online harmful information, as each can contribute to social, informational, or interpretive harm in online discourse, even though they differ in intent and mechanism. In this work, we use the term harmful content to jointly refer to misinformation-related harms and social and interpretive harms.
\begin{itemize}[leftmargin=*]

\item Misinformation-related harms:
    \begin{itemize}
    \item \textbf{\textit{Conspiracy}} — Content that promotes exaggerated claims such as stolen elections, secret political plots, or suppressed candidates. These narratives often spread rapidly and can distort political perceptions or encourage radicalization~\cite{douglas2017psychology, bartlett2010power}.\\
    
    \item \textbf{\textit{Sensationalism}} — Alarmist content featuring dramatic or misleading statements, warnings about political or geopolitical events. Sensationalism is known to drive engagement and contribute to the virality of harmful content~\cite{mourao2019fake, law2021sensationalist}.\\
    
    \item \textbf{\textit{Speculation}} — Bold assertions lacking credible evidence, often involving predictions or allegations about political outcomes or global events~\cite{blom2021potentials, rojecki2016rumors}.\\
    \end{itemize}
    
\item Social and interpretive harms:
    \begin{itemize}
    \item \textbf{\textit{Hate Speech}} —  Posts containing personal attacks, slurs, or inflammatory rhetoric targeting political figures or marginalized groups, which may incite violence or discourage civic engagement~\cite{rasaq2017media, solovev2022hate, banko2020unified}. We include this label because hate speech can overlap with discussions of polarized or misleading content~\cite{cinelli2021dynamics}, such as when false stereotypes are propagated, or specific groups are targeted. \\  
    
    \item \textbf{\textit{Satire}} — Political satire that uses humor, irony, or exaggeration to critique public figures, ideologies, or policies. While often protected as free expression, satire can blur the line between commentary and harmful content~\cite{kulkarni2017internet, bedard2018satire}.
    \end{itemize}
\end{itemize}

\subsection{AI Annotation (LLM)}
\label{sec:AI_annotation_method}
Our study setup treats each LLM as an independent annotator. For each post, we collect predictions from six LLMs, two from each model family.  
\begin{itemize}
    \item  \textbf{Lighter-weight Models}: ``gpt-4o-mini-2024-07-18'', ``Gemini-2.0-flash'', ``Llama-3.1-8B-Instruct''\\
    \item \textbf{Full Models}: ``gpt-4o-2024-08-06'', ``gemini-2.5-pro'', ``Llama-3.3-70B-Instruct''
\end{itemize}
These models were chosen based on their parameter scale and top benchmark performance on the Vellum.ai LLM Leaderboard~\footnote{https://www.vellum.ai/llm-leaderboard} at the time of the investigation. Selection criteria emphasized general performance (MMLU), coding efficiency (HumanEval), reasoning abilities (GPQA), scalability, and efficiency.
Larger models like GPT-4o and LLAMA 3.3 excel in MMLU and are suitable for complex tasks, while GPT-4o (90.2\%) performs exceptionally well on HumanEval for accurate zero-shot code generation. Lighter-weight models offer efficient alternatives for resource-limited scenarios in reasoning tasks, without significant performance trade-offs. 

We adopted a zero-shot learning approach, prompting each LLM to classify posts into predefined categories based on detailed definitions. We set each LLM's \textit{Temperature} to 0 to improve reproducibility and reduce response variability. For each post, the model returned a boolean value (\textit{True} or \textit{False}) 
for each category, indicating whether the category was present or absent. We enforced structured responses in the output to minimize formatting errors and repeated the annotation process when misformatted responses occurred. We designed the following prompt for our study: 

\begin{tcolorbox}[colback=gray!5!white, colframe=gray!75!black, title=LLM Prompt]
\label{prompt} 
\footnotesize
\textbf{Your task} is to accurately classify social media posts related to the U.S. Presidential Election.  Determine whether the given post falls into one or more of the following categories: \textbf{Conspiracy, Sensationalism, Hate Speech, Speculation, and Satire}.  Use the detailed definitions provided for each category and respond with \textbf{True} or \textbf{False} for each category only.  
\begin{itemize}[leftmargin=*]
    \item \textbf{Conspiracy:} \textit{Simplifies complex events by attributing them to secret plots, rejects mainstream information, forms closed belief communities, replaces science with alternative explanations, or frames events as elite deception.}
    \item \textbf{Sensationalism:} \textit{Uses exaggerated or dramatic language, shock and fear appeal, oversimplifies issues, or employs clickbait-style framing to increase engagement.}
    \item \textbf{Hate Speech:} \textit{Contains incitement of discrimination, defamation, hostility, or violence based on identity, or makes false statements that damage a person’s reputation (libel/slander).}
    \item \textbf{Speculation:} \textit{Circulates unverified claims for political advantage, driven by partisan interests, amplified in ideological echo chambers, and sustains political controversy.}
    \item \textbf{Satire:} \textit{Uses humor, political satire, and internet memes to criticize or comment on politics, often spreading through viral online platforms.}
\end{itemize}
\textbf{Post:} \texttt{"\{post\}"}
\end{tcolorbox}

\subsection{Human Annotation (MTurk)}
\label{human_annotation_method} 
To benchmark the performance between human and AI-generated annotations, we recruited 34 human annotators on Amazon Mechanical Turk (MTurk) \footnote{https://www.mturk.com/} in late September 2025. The study was approved by the Institutional Review Board of the affiliated university (Protocol \#2037012-3). Due to labor costs, we randomly sampled 1,000 posts (1.02\%).Each MTurk worker successfully labeled approximately 88 posts on average, with 5 annotations per post.
Although this is a small subset, simulation evidence suggests that when intercoder reliability is acceptable (>0.7), manually annotating about 1\% of the full dataset is sufficient to validate supervised classification models. This level of annotation yields reliable results with low false positive and false negative rates, even for subjective short-form social media text~\cite{song2020validations}.
In each session, the MTurk worker first provided informed consent and completed the same annotation tasks as the LLM, using identical instructions and category definitions (as given to the LLM in Section \ref{sec:AI_annotation_method}). 

Both LLMs and human annotators are presented with the same raw post, including hashtags and URLs. Only emojis (which appear in fewer than 10\% of posts) are excluded from the human annotation interface due to MTurk limitations.
MTurk workers also completed a brief demographic questionnaire including their political orientation. Each post was independently annotated by three unique workers. Workers were compensated at a fair rate equivalent to \$15 per hour. 
To ensure annotation quality, we applied strict qualification criteria following prior work \cite{zhang2023needle}. Each eligible worker was required to hold MTurk Master status, have completed more than 1,000 previously approved tasks, maintain a task approval rate above 95\%, and be located in the United States. The demographic composition of human annotators is shown in Table \ref{tab:demographics_compact_reordered}.

\begin{table}[!htb]
\centering
\footnotesize
\caption{Demographic Composition of Human Annotators (N = 34). AI experience refers to prior use of AI specifically for annotation or labeling tasks.}
\label{tab:demographics_compact_reordered}
\begin{tabular}{lrr|lrr}
\toprule
\textbf{Category} & \textbf{Count} & \textbf{\%} & \textbf{Category} & \textbf{Count} & \textbf{\%} \\
\midrule
\multicolumn{3}{c}{\textbf{Age}} & \multicolumn{3}{c}{\textbf{Gender}} \\
25--34 years & 2  & 5.9  & Male   & 24 & 70.6 \\
35--44 years & 15 & 44.1 & Female & 10 & 29.4 \\
45--54 years & 11 & 32.4 & Non-binary & 0 & 0 \\
55+ years    & 6  & 17.6 & Prefer not to say& 0&0 \\
\midrule
\multicolumn{3}{c}{\textbf{Annual Household Income}} & \multicolumn{3}{c}{\textbf{Area of Residence}} \\
Less than \$20k       & 6  & 17.6 & Rural         & 11 & 32.4 \\
\$20k–\$30k           & 4  & 11.8 & Suburban      & 11 & 32.4 \\
\$30k–\$40k           & 3  & 8.8  & Urban         & 7  & 20.6 \\
\$40k–\$50k           & 1  & 2.9  & Metropolitan  & 5  & 14.7 \\
\$50k–\$75k           & 10 & 29.4 & Other & 0 & 0 \\
\$75k–\$100k          & 5  & 14.7 & & & \\
\$100k–\$150k         & 4  & 11.8 & & & \\
More than \$150k       & 1  & 2.9  & & & \\
\midrule
\multicolumn{3}{c}{\textbf{Political Ideology}} & \multicolumn{3}{c}{\textbf{Political Affiliation}} \\
Very Liberal   & 5  & 14.7 & Democrat    & 14 & 41.2 \\
Liberal        & 11 & 32.4 & Republican  & 3  & 8.8  \\
Centrist       & 11 & 32.4 & Independent & 16 & 47.1 \\
Conservative   & 5  & 14.7 & Other: Socialist & 1 & 2.9 \\
Very Conservative & 1 & 2.9 & Prefer not to say & 0 & 0 \\
Prefer not to say & 1 & 2.9 & & & \\
\midrule
\multicolumn{3}{c}{\textbf{Education}} & \multicolumn{3}{c}{\textbf{AI Annotation Experience}} \\
High School / GED      & 4  & 11.8 & Strongly Disagree    & 24 & 70.6 \\
Some College           & 10 & 29.4 & Somewhat Disagree    & 4  & 11.8 \\
Bachelor's Degree      & 18 & 52.9 & Neutral              & 1  & 2.9 \\
Master's Degree        & 1  & 2.9  & Somewhat Agree       & 2  & 5.9 \\
Doctorate or Higher        & 1  & 2.9  & Strongly Agree       & 3  & 8.8 \\
\bottomrule
\end{tabular}
\end{table}

\subsection{Inter-rater Reliability (IRR)}
To assess Inter-rater Reliability (IRR), we use the Average Pairwise Cohen’s Kappa ($\kappa$) \cite{warrens2011cohen}, Average Pairwise Percentage Agreement, and Krippendorff’s Alpha ($\alpha$) \cite{krippendorff2018content} for each category. These measurements are commonly established and follow guidelines suggested by \citet{10.1145/3359174}.
First, we evaluate IRR across LLMs to assess agreement among AI models and identify consistency patterns. Second, we compute IRR among human annotators to examine the variation of human-labeled posts. Third, we compare LLM annotations with human annotations. We derive consensus human labels using majority voting and treat them as ground truth. We then compared this against individual LLM performance, the majority vote of three LLMs, and the majority vote of five LLMs, covering all 32 possible combinations. We report Precision, Recall, and F1-score for all combinations to evaluate the performance of using LLMs as a supportive annotation tool.

\begin{table*}[!htb]
\caption{Cost--benefit comparison of LLM models and human annotation, based on classifying a total of 1,000 social media posts with five binary annotations each during the experiment. Asterisk (*) denotes the estimated value.}
\footnotesize
\centering
\begin{tabular}{p{2cm}>{\centering\arraybackslash}p{1.6cm}p{1.8cm}p{2.4cm}p{2.6cm}p{2.2cm}}
\toprule
Model & Open Source & Parameters (n) & Knowledge Cut-off & Computation Time (Avg) & Cost (Avg) \\ 
\midrule
GPT--4o Mini & $\times$ & 8 B* & Oct 2023 & 100 sec & \$0.09 \\
GPT--4o & $\times$  & 200 B* & Oct 2023 & 88 sec & \$1.43 \\ 
Gemini 2.0 Flash & $\times$ & 20 B* & Aug 2024 & 60 sec & \$0.05 \\ 
Gemini 2.5 Pro & $\times$  & 140 B* & Jan 2025 & 30 min & \$8.97 \\ 
Llama 3.1 & \checkmark & 8 B & Dec 2023 & 130 sec & \$0.17 (4\texttimes{} Nvidia L4) \\ 
Llama 3.3 & \checkmark  & 70 B & Dec 2023 & 196 sec & \$0.97 (4\texttimes{} Nvidia A100) \\ 
\midrule
\multicolumn{3}{c}{Single Crowdsource Worker} &  & 7.78 hr* & \$130* \\
\bottomrule  
\end{tabular}
\label{tab:model_comparison}
\end{table*}

\begin{figure*}[!htb]
    \centering
    \includegraphics[trim={0 0.2cm 0 0.4cm},clip, width=0.9\textwidth]{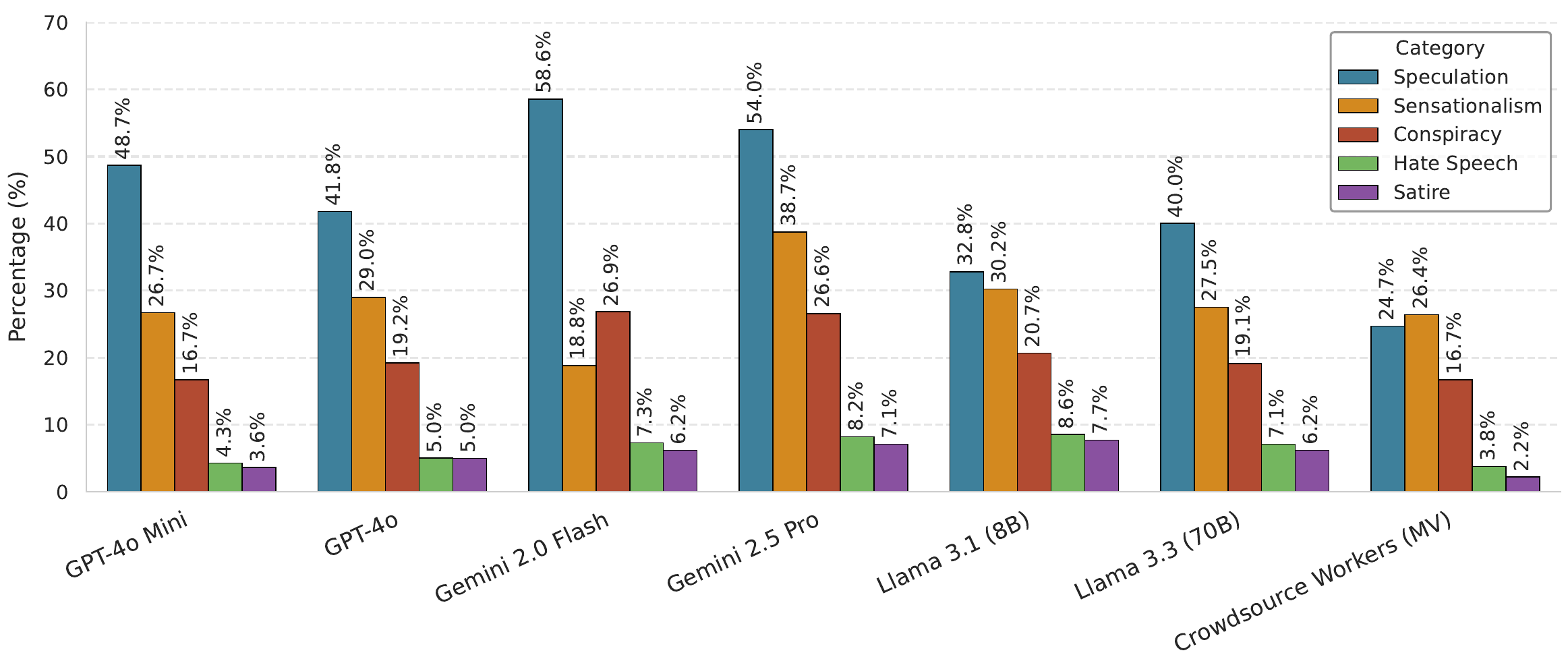}
    \caption{Distribution of each category annotated by different LLM and crowdsourced workers (majority vote). Each category was evaluated independently.}
    \label{fig:distribution}
\end{figure*}

\section{Results}
We first present LLM agreement patterns, followed by inter-annotator agreement from MTurk. We then compare LLM and human annotations to assess their alignment, discrepancies, and sources of disagreement.

\subsection{LLM Agreement}

\subsubsection{Cost-benefit Comparison}
Table \ref{tab:model_comparison} presents a comparative analysis of AI models, examining open-source status, parameter counts, knowledge cutoffs, computational times, and associated costs, highlighting the trade-offs between model size, performance, and operational expenses.
GPT model parameter counts are estimated from~\cite{abacha2025medec}, while Gemini model parameters are estimated from public sources \footnote{\url{https://www.emergentmind.com/topics/google-gemini-2-0-flash}}.
Most models in our study had a knowledge cutoff prior to curation of the dataset from X, minimizing the risk of data contamination. Gemini 2.5 Pro, with a cutoff in January 2025, may have had minimal exposure to the dataset. That being said, we included it for completeness, as it represents one of the full models from the Gemini family, since the Gemini 1.5 Pro was deprecated during our investigation. This setup allows a fair evaluation of LLM performance while accounting for later models.
Gemini 2.5 Pro is the most recent LLM model included in this study and costs more at the time of the experiment. We expect its cost to decrease over time.
The lighter models tend to run faster and are less expensive, while the larger models are more costly and take more time. The human annotation estimate was calculated based on MTurk’s average completion time (28 seconds per post) at a rate of \$15 per hour.

\subsubsection{Distribution Comparison}
Figure \ref{fig:distribution} compares six LLMs and human annotation (majority vote) based on annotation counts across five categories and their distributions.
Results indicate that LLMs generally detect a high proportion of post content as Speculation (ranging from 32.8\% - 58.6\%), Sensationalism (ranging from 18.8\% - 38.7\%), and Conspiracy (ranging from 16.7\% - 26.9\%). On the other hand, the lower categories identified by LLMs are generally Hate Speech (ranging from 4.3\% - 8.6\%) and Satire (ranging from 3.6\% - 7.7\%). These distribution patterns are similar to those found in human annotations. Gemini 2.5 Pro consistently has a higher label count across nearly all categories compared to the other models.

\subsubsection{IRR Among LLMs}
Using all combinations of the six available LLM annotations, we first computed 15 pairwise IRR scores using percentage agreement for each category
(Table~\ref{tab:agreement} in the Appendix). Highly distributed categories, such as Conspiracy ($91.01\% \pm 1.76$), Speculation ($78.82\% \pm 5.24$), and Sensationalism ($82.83\%\pm2.72$), demonstrated substantial agreement. Less frequent categories, like Hate Speech ($94.75 \%\pm 1.30$) and Satire ($94.22 \%\pm 1.39$), achieved even higher agreement. Because percentage agreement only measures the proportion of matching labels, it does not account for chance or rater variability.
To address this, we calculated pairwise Cohen’s Kappa ($\kappa$).
Conspiracy exhibited the highest agreement ($\kappa = 0.75 \pm 0.05$), followed by Hate Speech ($\kappa = 0.62 \pm 0.08$), Sensationalism ($\kappa = 0.60 \pm 0.07$), Speculation ($\kappa = 0.58 \pm 0.09$), and Satire ($\kappa = 0.53 \pm 0.09$), indicating a moderate to strong level of agreement overall, with the best performing pair achieving strong agreement as defined by \cite{mchugh2012interrater}. Llama and Gemini have the lowest $\kappa$ (for example, Llama 3.1 8B and Gemini 2.5 Pro have $\kappa$ of 0.66 in Conspiracy), and this pattern is consistent across all categories.

Finally, we computed Krippendorff’s Alpha ($\alpha$) for all sets of three LLMs, with results presented in Figure \ref{fig:IRR_LLM_krippendorff}. This metric captures agreement across multiple raters while accounting for chance and variation, providing a robust measure of reliability. The results are consistent with the pairwise $\kappa$ values: Conspiracy shows the highest agreement, and Satire shows the lowest within LLMs.
These results show that most labels achieved moderate to strong agreement, and while Llama and Gemini scored slightly lower, this highlights opportunities to improve their consistency, especially for sensitive labels like Conspiracy, through better tuning or calibration.

\subsection{Human Annotation Agreement}
\subsubsection{Demographic Variation}
\label{Demographic_Variation}
\begin{figure}[!hbtp]
  \centering
  \includegraphics[width=0.99\linewidth]{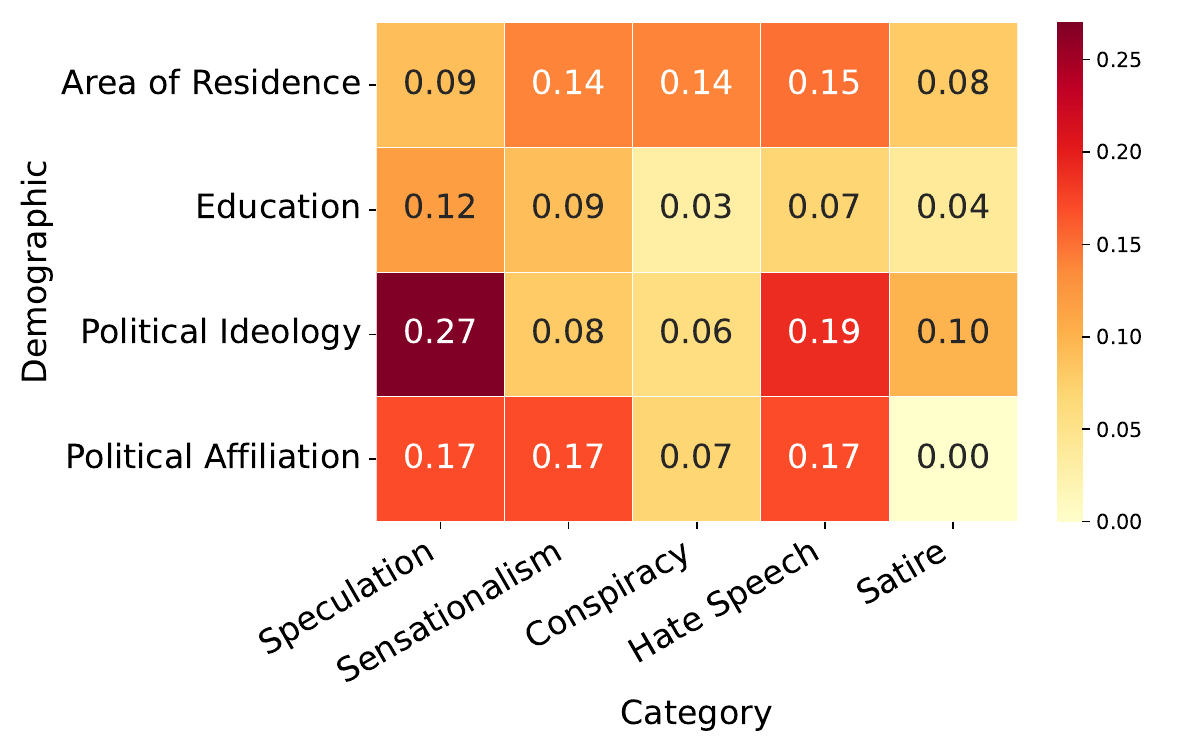}
  \caption{Strength of association between annotators’ demographics and annotation categories, expressed as Cramér's~V from Pearson's~$\chi^{2}$ tests ($N\approx3{,}000$ assignments).  
  Darker cells indicate stronger associations.}
  \label{fig:chi2_heatmap}
\end{figure}

Pearson's~$\chi^{2}$ analyses (Figure~\ref{fig:chi2_heatmap}) revealed statistically detectable but generally small associations between demographics and labeling patterns. 
Among the five labels, \textit{Speculation} showed the strongest demographic influence, driven largely by political ideology ($\chi^{2}_{(4)}=222.49,\,p<0.001,\,V=0.27$) and to a lesser extent by political affiliation ($\chi^{2}_{(3)}=89.80,\,p<0.001,\,V=0.17$).  
\textit{Hate~Speech} was the next most variable across groups, differing by both ideology ($\chi^{2}_{(4)}=111.99,\,p<0.001,\,V=0.19$) and political affiliation ($\chi^{2}_{(3)}=85.76,\,p<0.001,\,V=0.17$).  

\begin{figure*}[!htb]
    \centering
    \begin{subfigure}{0.49\linewidth}
        \centering
        \includegraphics[width=0.9\linewidth]{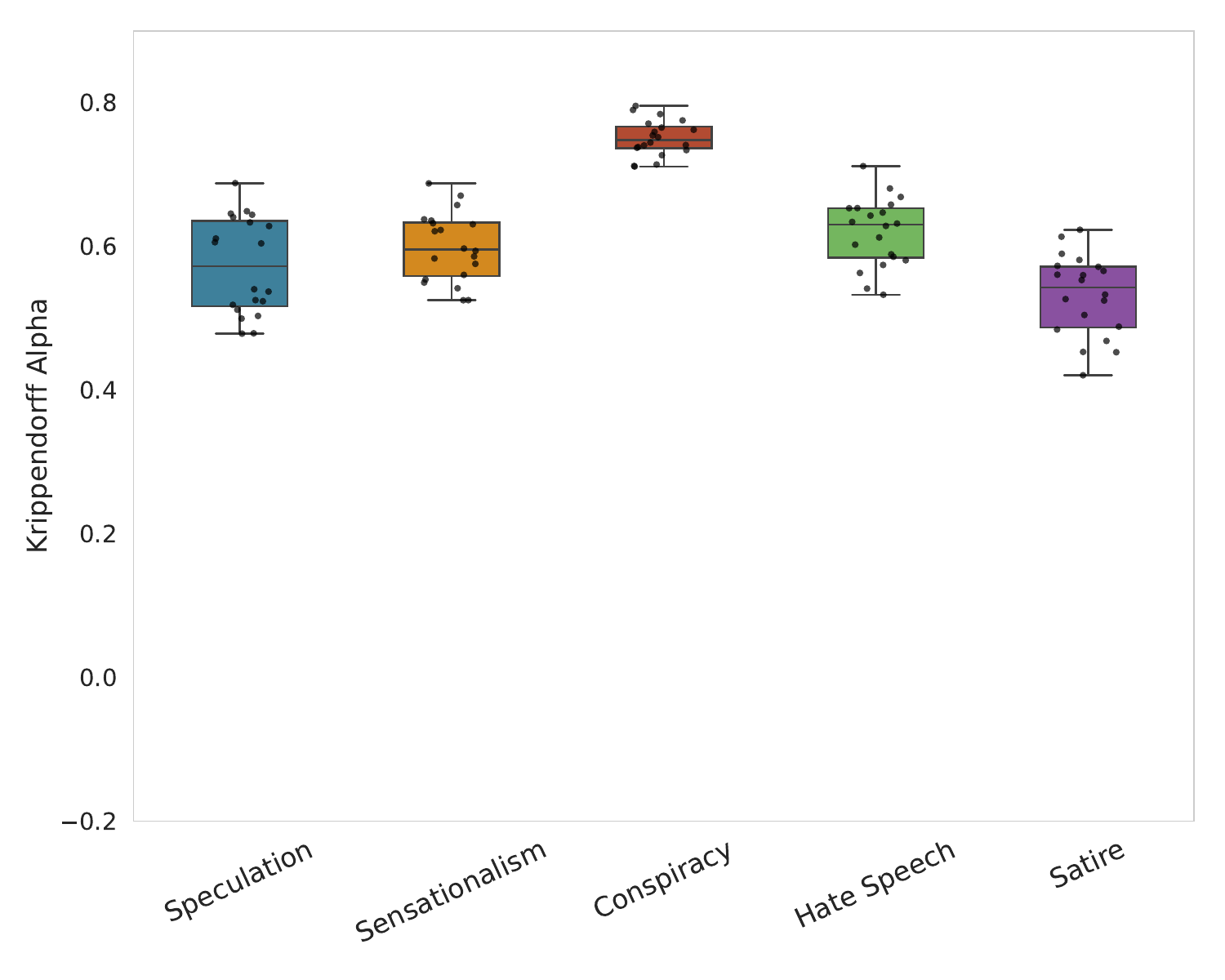}
        \caption{LLMs (n=20)}
        \label{fig:IRR_LLM_krippendorff}
    \end{subfigure}
    \begin{subfigure}{0.49\linewidth}
        \centering
        \includegraphics[width=0.9\linewidth]{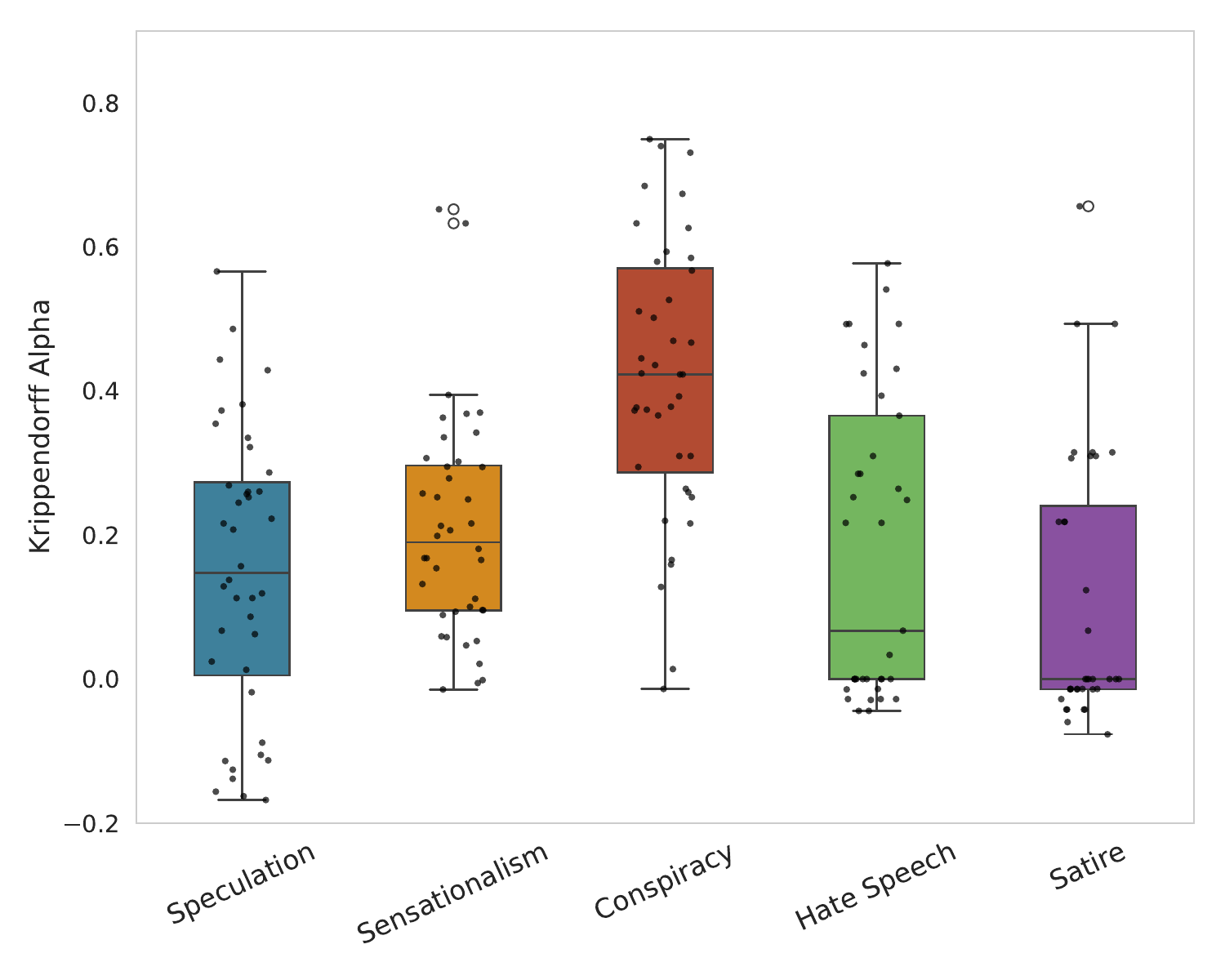}
        \caption{MTurk Workers (n=193)}
        \label{fig:IRR_AMT_krippendorff_alpha}
    \end{subfigure}
    \caption{Krippendorff’s Alpha ($\alpha$) Comparison. (a) shows $\alpha$ computed for all combinations of three LLMs selected from six models across the full dataset. (b) presents $\alpha$ for 193 groups of human annotators on the subset of data where annotations were available. The values range from -1 to 1, where higher $\alpha$ values indicate stronger agreement.}
    \label{fig:IRR_krippendorff_combined}
\end{figure*}

Area of residence showed small but significant effects across all labels ($V \approx 0.08$--$0.15$), while education was significant for \textit{Hate~Speech}, \textit{Sensationalism}, and \textit{Speculation} ($p < 0.001,\; V \approx 0.07$--$0.13$) but not for \textit{Conspiracy} or \textit{Satire}. For political affiliation, posts labeled as \textit{Conspiracy} were more common among certain political groups, showing a significant but weak association ($\chi^{2}_{(3)}=17.28,\,p<0.001,\,V=0.07$), while \textit{Satire} showed no notable differences across groups. Ideology also influenced labeling across all categories: \textit{Conspiracy} showed only a mild association ($\chi^{2}_{(4)}=15.54,\,p<0.01,\,V=0.06$), whereas \textit{Speculation} was labeled far more often by some ideological groups ($\chi^{2}_{(4)}=222.49,\,p<0.001,\,V=0.27$), making it the strongest group-level difference observed. Trend tests further indicated clear monotonic shifts across the liberal$\to$conservative spectrum: annotators further to the right were more likely to label content as \textit{Hate~Speech} ($p<0.001$, $\rho\approx0.14$), while they were slightly less likely to label content as \textit{Speculation} ($p<0.001$, $\rho\approx-0.07$). Overall, trends were statistically reliable but small in size. 


\begin{table*}[!htb] 
\centering
\footnotesize

\caption{Comparison of Cohen’s Kappa ($\kappa$) across individual LLM models and the LLM majority vote (MV) from three LLMs, evaluated against crowdsourced workers' consensus labels based on MV. The smallest value in each category is highlighted in \colorbox{red!20}{red}, and the largest value is highlighted in \colorbox{green!20}{green}. Mean ± SD and range are reported based on the 15 combination groups.}
\label{tab:llm_AMT_kappa_compare_1000post}
\begin{tabular}{p{1.65cm}cc|cc|cc|cc|cc}
\toprule
\textbf{Model} & \multicolumn{2}{c}{\textbf{Speculation}} & \multicolumn{2}{c}{\textbf{Sensationalism}} & \multicolumn{2}{c}{\textbf{Conspiracy}} & \multicolumn{2}{c}{\textbf{Hate Speech}} & \multicolumn{2}{c}{\textbf{Satire}} \\
\midrule
GPT-4o Mini & \multicolumn{2}{c}{0.27} & \multicolumn{2}{c}{0.24} & \multicolumn{2}{c}{0.60} & \multicolumn{2}{c}{0.36} & \multicolumn{2}{c}{0.24} \\

GPT-4o & \multicolumn{2}{c}{0.34} & \multicolumn{2}{c}{\colorbox{red!20}{0.17}} & \multicolumn{2}{c}{0.62} & \multicolumn{2}{c}{0.46} & \multicolumn{2}{c}{\colorbox{red!20}{0.17}}
\\
Gemini 2.0 Flash & \multicolumn{2}{c}{\colorbox{red!20}{0.26}} & \multicolumn{2}{c}{0.23} & \multicolumn{2}{c}{0.60} & \multicolumn{2}{c}{0.43} & \multicolumn{2}{c}{0.23} \\
Gemini 2.5 Pro & \multicolumn{2}{c}{0.28} & \multicolumn{2}{c}{\colorbox{red!20}{0.17}} & \multicolumn{2}{c}{\colorbox{red!20}{0.56}} & \multicolumn{2}{c}{\colorbox{red!20}{0.28}} & \multicolumn{2}{c}{\colorbox{red!20}{0.17}} \\
Llama 3.1 (8B) & \multicolumn{2}{c}{\colorbox{green!20}{0.37}} &\multicolumn{2}{c}{0.26} & \multicolumn{2}{c}{0.60} & \multicolumn{2}{c}{0.38} & \multicolumn{2}{c}{0.26} \\
Llama 3.3 (70B) & \multicolumn{2}{c}{0.28} & \multicolumn{2}{c}{0.20} & \multicolumn{2}{c}{0.59} & \multicolumn{2}{c}{0.39} & \multicolumn{2}{c}{0.20} \\
\midrule
\multirow{2}{*}{\textbf{LLMs MV}} & \textbf{Mean ± SD} & \textbf{Min - Max} & \textbf{Mean ± SD} & \textbf{Min - Max} & \textbf{Mean ± SD} & \textbf{Min - Max} & \textbf{Mean ± SD} & \textbf{Min - Max} & \textbf{Mean ± SD} & \textbf{Min - Max}  \\
 & 0.31 ± 0.03 & 0.27 – 0.36 & 0.31 ± 0.03 & 0.35 – \colorbox{green!20}{0.45}&   0.62 ± 0.01 & 0.60 – \colorbox{green!20}{0.65} & 0.43 ± 0.02 & 0.39 - \colorbox{green!20}{0.48}  & 0.23 ± 0.02 & 0.19 – \colorbox{green!20}{0.27} \\
  Best Set & \multicolumn{2}{c}{Llama3.1, GPT-4o, Gemi-2.5} & \multicolumn{2}{c}{Llama3.3, GPT-4o, Gemi-2.5} & \multicolumn{2}{c}{Llama3.1, GPT-4o, Gemi-2.0} & \multicolumn{2}{c}{Llama3.1, GPT-4o, Gemi-2.0} & \multicolumn{2}{c}{Llama3.1, Llama3.3, Gemi-2.5} \\
 \bottomrule
\end{tabular}
\end{table*}

The group-level mean-proportion analysis further illustrates how labeling patterns vary across demographics. \textit{Speculation} stood out as the most frequent label among Centrists and Independents, while \textit{Sensationalism} was relatively higher among Republicans and rural annotators. 
\textit{Hate~Speech} labeling was elevated among Republicans compared to other political groups, whereas \textit{Satire} remained rare and low across nearly all demographics.  
Political affiliation overall predicted higher odds of \textit{Conspiracy}, \textit{Sensationalism}, and \textit{Speculation} among Republicans and Independents, while ideology showed both positive and negative gradients depending on the label (\emph{e.g.,} conservatives tended to apply fewer \textit{Speculation} tags but more \textit{Hate~Speech}). 
See Figures~\ref{fig:means_by_demographics} in the Appendix for additional plots of these group-level mean values.

These demographic patterns help explain the distributional differences we observe across categories and their agreement scores. 
For example, the \textit{Speculation} label, which is strongly shaped by ideology and political affiliation, is the most frequently selected category by both LLMs and crowd workers (Figure \ref{fig:distribution}) and exhibits relatively high inter rater agreement (Figures \ref{fig:IRR_LLM_krippendorff} and \ref{fig:IRR_AMT_krippendorff_alpha}).
By contrast, \textit{Satire} was consistently rare across models and workers and exhibited the lowest agreement, underscoring its ambiguous boundaries. 
These complementary findings suggest that systematic demographic preferences subtly influence how people label news content, shaping category frequencies and partially explaining variation in model human agreement.

\begin{figure*}[!htbp]
    \centering
    \includegraphics[trim=0 0.5cm 0 0, clip, width=0.85\linewidth]{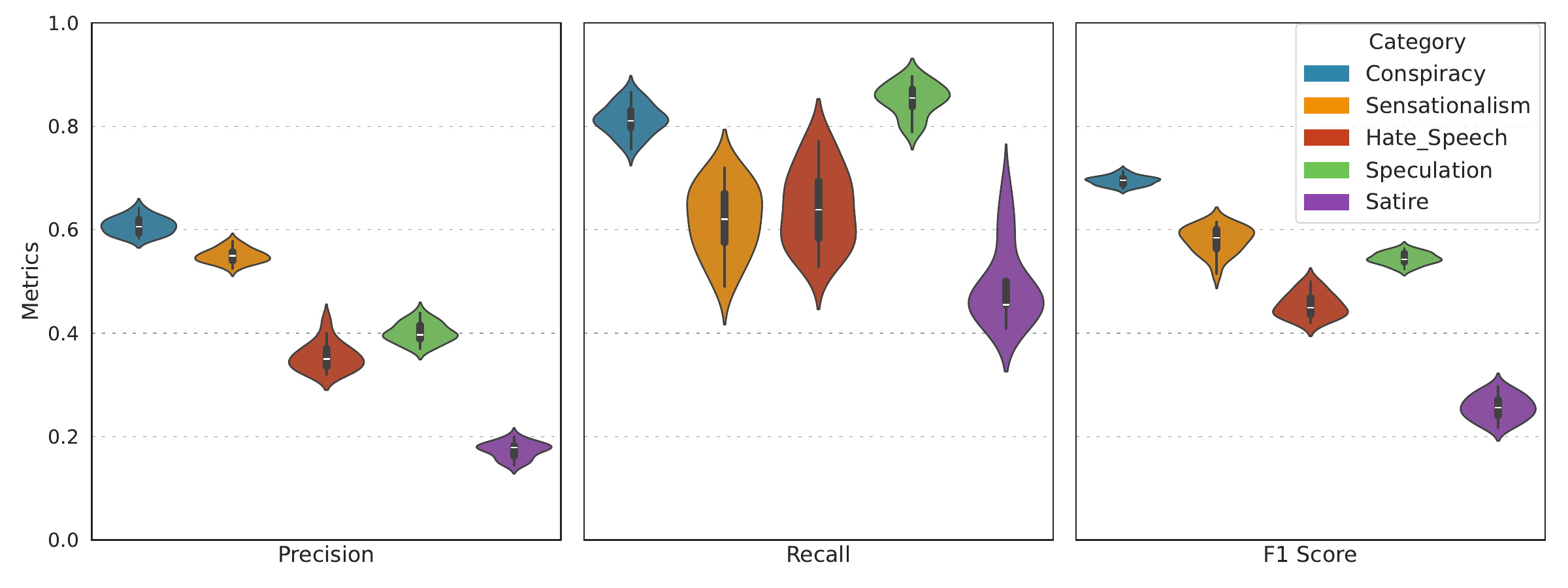}
    \caption{Direct comparison between majority-vote labels produced by LLM ensembles (20 outputs) and majority-vote labels from crowdsourced workers across different categories.}
    \label{fig:F1_Recall_precision_LLM}
\end{figure*}

\subsubsection{IRR Among Human Annotators}
Each batch of annotation tasks included 25 posts, with a total of 34 unique MTurk workers, resulting in 193 complete sets for IRR comparison. Figure \ref{fig:IRR_krippendorff_combined} shows Krippendorff’s $\alpha$, and Appendix \ref{tab:agreement} shows the pairwise percentage agreement and the pairwise Cohen’s Kappa ($\kappa$).
For Conspiracy, the pairwise $\kappa$ was 0.43 (range 0.08–0.75) with an average agreement of 82.94\% (range 68.0–94.67\%). Hate Speech exhibited moderate agreement (mean $\kappa = 0.32$, mean agreement 91.34\%, $\alpha = 0.18$), while Sensationalism showed lower reliability (mean $\kappa = 0.23$, mean agreement 68.33\%, $\alpha = 0.21$). Speculation displayed moderate variability (mean $\kappa = 0.21$, $\alpha = 0.15$), and Satire had high agreement percentages despite low $\kappa$ (mean $\kappa = 0.16$, mean agreement 92.06\%, $\alpha = 0.11$). 
Overall, frequent categories such as Conspiracy show higher agreement, while more subjective categories like Satire exhibit greater variability.

\subsection{LLM vs. Human Annotation Comparison}
\subsubsection{Human Consensus Labels} 
To ensure label quality in subjective annotation tasks, employing a majority vote method is a common practice. This approach leverages \textit{the Wisdom of Crowds} to mitigate individual biases and errors, leading to more accurate and consistent labels. This technique is particularly effective when individual annotators have varying levels of expertise or when tasks are subjective.
For instance, a previous study highlights that majority voting is the simplest efficient consensus algorithm, resulting in high-quality integrated labels, provided that the annotators are reasonably reliable \cite{ZHANG2017254}.   Similarly, another empirical study provides a theoretical analysis demonstrating that the label error rate decreases exponentially with the number of workers selected for each task, underscoring the effectiveness of majority voting in improving label quality \cite{wang2015crowdsourcing}. Therefore, we applied majority voting across three human annotators for each post and category to derive the ground-truth labels. The distribution of majority vote labels is shown in Figure \ref{fig:distribution} (see bars for Crowdsource Workers (MV)), which exhibits a pattern similar to that of the LLM annotations, with Hate Speech and Satire appearing less frequent and the remaining categories occurring more commonly.

\subsubsection{LLM Performance}
With the potential of using LLMs as supporting annotation tools, we first computed Cohen’s kappa ($\kappa$) for each LLM against the human consensus labels. The results are presented in Table \ref{tab:llm_AMT_kappa_compare_1000post}. Llama 3.1 (8B) generally performs better across most categories, while Gemini 2.5 Pro shows slightly lower performance.
Similar to the human-majority-vote approach, we also applied it across multiple LLMs using the same method. Results, also shown in Table \ref{tab:llm_AMT_kappa_compare_1000post}, indicate that LLM majority voting generally improves agreement, particularly when combining models from different LLM families. An exception is the Speculation category, where a single Llama 3.1 (8B) model achieves the higher agreement score. In addition, comparing the majority vote of five LLMs did not improve IRR and, in fact, led to slightly lower agreement.

To further assess direct performance, we report recall, precision, and F1 score in Figure~\ref{fig:F1_Recall_precision_LLM}, examining how LLM consensus predictions align with human consensus labels. Since annotations in misinformation tasks are often used for detection and typically focus on the positive class, Recall can serve as a more representative metric. The LLMs achieved the highest Recall in Speculation (mean = 0.85 $\pm$ 0.03) and Conspiracy (mean = 0.81 $\pm$ 0.03), followed by Sensationalism (mean = 0.62 $\pm$ 0.07) and Hate Speech (mean = 0.64 $\pm$ 0.08). Satire showed the lowest Recall (mean = 0.49 $\pm$ 0.08). These results indicate that LLM majority voting aligns well with human consensus for clearly defined categories, while subjective or nuanced categories like Satire remain more challenging.

\section{Discussion}
Across all evaluated models, IRR among LLMs is consistently higher than that observed among human annotators, indicating more systematic and stable labeling behavior. In contrast, human annotators showed lower and more variable agreement overall, especially in subjective categories such as Satire.
Therefore, LLMs show considerable potential as reliable support tools for multi-label annotation tasks, particularly when combining models from different families. These ‘out-of-the-box’ models perform competitively in complex online harmful content detection without fine-tuning. Aggregating predictions via majority voting across diverse LLMs further enhances robustness, surpassing single-model performance and human inter-annotator agreement. This approach reduces labeling costs while maintaining performance comparable to human annotation, demonstrating its practical value for large-scale analysis of harmful content.
Using a wisdom-of-the-crowd approach across the best set in Table \ref{tab:llm_AMT_kappa_compare_1000post}, we curate a robust multi-label dataset. In the final dataset, 59.62\% of posts receive at least one harmful content label, indicating that potentially harmful content is widespread in online discussions rather than a rare edge case.

Our analysis also confirms that demographics measurably influenced how MTurk workers applied the five content labels. Speculation was the most ideology-sensitive label applied most often by Centrists and Independents, while conservatives labeled fewer posts as Speculation but more as Hate Speech. This shows that crowd-sourced “ground-truth” labels are not demographically neutral and that annotators’ worldviews shape both category frequencies and model–human agreement. In contrast, Satire remained rare across all groups and showed low agreement, reflecting its inherently ambiguous boundaries rather than demographic bias. These results also set a realistic baseline for human annotation quality.

Human agreement reached the highest Krippendorff's $\alpha$ among the Conspiracy.
As discussed in Section \ref{human_annotation_method}, enforcing strict selection criteria helped improve the quality of annotations. Initially, without such criteria, workers often neglected annotation tasks, leading to patterns such as selecting all true or all false responses. After applying more rigorous qualification standards, this behavior was no longer observed. With additional training and practice, the IRR of human annotators could likely be further improved. 
Similar strategies—such as one-shot, few-shot, or fine-tuning could also improve LLM performance with minimal overhead, and future work should explore these to boost both human and LLM annotation consistency and efficiency.

\section{Limitations}
Although the dataset could be expanded, X.com API access is limited and costly. The 100k collected posts remain a meaningful contribution, and future work could explore other social media platforms.
The terminology and labeling scheme used here differ from traditional fact-checking labels such as \textit{factual}, \textit{partially factual}, \textit{fake}, \textit{accurate}, or \textit{partially accurate}. These labels are common in fact-checking and news credibility research, where the goal is to assess content based on truthfulness or accuracy. Therefore, direct comparison with existing datasets or pre-trained models that use such labels is not possible. Our current framework in this study also fails to distinguish between bot- and human-generated content, despite their distinct linguistic and behavioral patterns. Future work integrating bot detection techniques could help refine the analysis of online harmful content dynamics across user types. In addition, occasional missing predictions occurred due to model overload (\emph{e.g.,} Gemini 2.5 Pro rate limits), which were mitigated by re-running affected samples. Notably, Gemini Flash exhibited a higher number of missing responses for Hate Speech (n = 3989, 4\%) and Satire (n = 3970, 4\%), while other categories remained unaffected.  
Although most models (5/6) have knowledge cutoffs preceding our dataset, LLMs may still reflect overlapping or biased training data, complicating comparisons with human annotators. Future work could explore LLM-based personas aligned with annotator demographics.

\section{Conclusion}
This study presents a large-scale, rigorously annotated dataset of U.S. election-related social media posts for harmful content research, evaluated through a comprehensive crowdsourced study. Leveraging consensus techniques and human oversight, we provide scalable labels that improve upon traditional manual or heuristic approaches. 
Our approach offers a practical baseline with a recall reaching up to 0.9, balancing quality and scalability for very large corpora. Majority voting across multiple LLMs mirrors human protocols while greatly reducing time and cost—critical for analyzing the billions of posts generated daily on online social networks. 
We also show that annotator demographics influence labeling tendencies, highlighting the need to consider human factors alongside model calibration in future pipelines. The dataset further supports exploration of co-occurring online harmful content signals, such as hate speech with satire, and is publicly available at: \url{https://github.com/Sensify-Lab/USE24-XD}.

\bibliographystyle{ACM-Reference-Format}
\bibliography{reference}

@article{pinto2024tracking,
  title={Tracking the 2024 US Presidential Election Chatter on Tiktok: A Public Multimodal Dataset},
  author={Pinto, Gabriela and Bickham, Charles and Salkar, Tanishq and Luceri, Luca and Ferrara, Emilio},
  journal={arXiv preprint arXiv:2407.01471},
  year={2024}
}

@article{raza2024fakewatch,
  title={FakeWatch: a framework for detecting fake news to ensure credible elections},
  author={Raza, Shaina and Khan, Tahniat and Chatrath, Veronica and Paulen-Patterson, Drai and Rahman, Mizanur and Bamgbose, Oluwanifemi},
  journal={Social Network Analysis and Mining},
  volume={14},
  number={1},
  pages={142},
  year={2024},
  publisher={Springer}
}

@article{al2024harmful,
  title={What is harmful content?},
  author={Al-Khoury, Dareen},
  journal={The Routledge Handbook of Conflict and Peace Communication},
  year={2024},
  publisher={Taylor \& Francis}
}

@article{zeng2023misinformation,
  title={The Misinformation},
  author={Zeng, Jing and Brennen, Scott Babwah},
  journal={Internet Policy Review},
  volume={12},
  number={4},
  pages={1725},
  year={2023},
  publisher={Alexander von Humboldt Institute for Internet and Society}
}

@book{wardle2017information,
  title={Information disorder: Toward an interdisciplinary framework for research and policymaking},
  author={Wardle, Claire and Derakhshan, Hossein},
  volume={27},
  year={2017},
  publisher={Council of Europe Strasbourg}
}

@article{tercova2025digital,
  title={Digital Skills' Role in Intended and Unintended Exposure to Harmful Online Content Among European Adolescents},
  author={Tercova, Natalie and Smahel, David},
  journal={Media and Communication},
  volume={13},
  year={2025},
  publisher={PRT}
}

@inproceedings{elsayed2019proposed,
  title={A proposed framework for improving analysis of big unstructured data in social media},
  author={Elsayed, Mohamed and Abdelwahab, Amira and Ahdelkader, Hatem},
  booktitle={2019 14th International conference on computer engineering and systems (ICCES)},
  pages={61--65},
  year={2019},
  organization={IEEE}
}

@article{velkova2023unstructured,
  title={Unstructured social media data processing with artificial intelligence},
  author={Velkova, Ivona},
  journal={Industry 4.0},
  volume={8},
  number={2},
  pages={65--67},
  year={2023},
  publisher={Scientific Technical Union of Mechanical Engineering" Industry 4.0"}
}

@inproceedings{banko2020unified,
  title={A unified taxonomy of harmful content},
  author={Banko, Michele and MacKeen, Brendon and Ray, Laurie},
  booktitle={Proceedings of the fourth workshop on online abuse and harms},
  pages={125--137},
  year={2020}
}

@article{harm,
author = {Arora, Arnav and Nakov, Preslav and Hardalov, Momchil and Sarwar, Sheikh Muhammad and Nayak, Vibha and Dinkov, Yoan and Zlatkova, Dimitrina and Dent, Kyle and Bhatawdekar, Ameya and Bouchard, Guillaume and Augenstein, Isabelle},
title = {Detecting Harmful Content on Online Platforms: What Platforms Need vs. Where Research Efforts Go},
year = {2023},
issue_date = {March 2024},
publisher = {Association for Computing Machinery},
address = {New York, NY, USA},
volume = {56},
number = {3},
issn = {0360-0300},
url = {https://doi.org/10.1145/3603399},
doi = {10.1145/3603399},
journal = {ACM Comput. Surv.},
month = oct,
articleno = {72},
numpages = {17}
}

@inproceedings{giachanou2020battle,
  title={The battle against online harmful information: The cases of fake news and hate speech},
  author={Giachanou, Anastasia and Rosso, Paolo},
  booktitle={Proceedings of the 29th ACM International Conference on Information \& Knowledge Management},
  pages={3503--3504},
  year={2020}
}

@inproceedings{choi2024automated,
  title={Automated claim matching with large language models: empowering fact-checkers in the fight against misinformation},
  author={Choi, Eun Cheol and Ferrara, Emilio},
  booktitle={Companion Proceedings of the ACM Web Conference 2024},
  pages={1441--1449},
  year={2024}
}

@inproceedings{nguyen2024human,
  title={Human vs ChatGPT: Effect of data annotation in interpretable crisis-related microblog classification},
  author={Nguyen, Thi Huyen and Rudra, Koustav},
  booktitle={Proceedings of the ACM Web Conference 2024},
  pages={4534--4543},
  year={2024}
}

@inproceedings{hamilton2024gpt,
  title={GPT Assisted Annotation of Rhetorical and Linguistic Features for Interpretable Propaganda Technique Detection in News Text.},
  author={Hamilton, Kyle and Longo, Luca and Bozic, Bojan},
  booktitle={Companion Proceedings of the ACM Web Conference 2024},
  pages={1431--1440},
  year={2024}
}

@inproceedings{nakamura2020fakeddit,
  title={Fakeddit: A new multimodal benchmark dataset for fine-grained fake news detection},
  author={Nakamura, Kai and Levy, Sharon and Wang, William Yang},
  booktitle={Proceedings of the twelfth language resources and evaluation conference},
  pages={6149--6157},
  year={2020}
}

@article{douglas2017psychology,
  title={The psychology of conspiracy theories},
  author={Douglas, Karen M and Sutton, Robbie M and Cichocka, Aleksandra},
  journal={Current directions in psychological science},
  volume={26},
  number={6},
  pages={538--542},
  year={2017},
  publisher={Sage Publications Sage CA: Los Angeles, CA}
}

@book{bartlett2010power,
  title={The power of unreason: Conspiracy theories, extremism and counter-terrorism},
  author={Bartlett, Jamie and Miller, Carl},
  year={2010},
  publisher={Demos London}
}

@article{mourao2019fake,
  title={Fake news as discursive integration: An analysis of sites that publish false, misleading, hyperpartisan and sensational information},
  author={Mour{\~a}o, Rachel R and Robertson, Craig T},
  journal={Journalism studies},
  volume={20},
  number={14},
  pages={2077--2095},
  year={2019},
  publisher={Taylor \& Francis}
}

@article{rasaq2017media,
  title={Media, politics, and hate speech: A critical discourse analysis},
  author={Rasaq, Adisa and Udende, Patrick and Ibrahim, Abubakar and Oba, La’aro},
  journal={E-Academia Journal},
  volume={6},
  number={1},
  year={2017}
}

@inproceedings{solovev2022hate,
  title={Hate speech in the political discourse on social media: Disparities across parties, gender, and ethnicity},
  author={Solovev, Kirill and Pr{\"o}llochs, Nicolas},
  booktitle={Proceedings of the ACM web conference 2022},
  pages={3656--3661},
  year={2022}
}

@article{blom2021potentials,
  title={The potentials and pitfalls of interactional speculations by journalists and experts in the media: The case of Covid-19},
  author={Blom, Jonas Nygaard and R{\o}nlev, Rasmus and Hansen, Kenneth Reinecke and Ljungdalh, Anders Kruse},
  journal={Journalism Studies},
  volume={22},
  number={9},
  pages={1142--1160},
  year={2021},
  publisher={Taylor \& Francis}
}

@article{rojecki2016rumors,
  title={Rumors and factitious informational blends: The role of the web in speculative politics},
  author={Rojecki, Andrew and Meraz, Sharon},
  journal={New Media \& Society},
  volume={18},
  number={1},
  pages={25--43},
  year={2016},
  publisher={Sage Publications Sage UK: London, England}
}

@article{kulkarni2017internet,
  title={Internet meme and Political Discourse: A study on the impact of internet meme as a tool in communicating political satire},
  author={Kulkarni, Anushka},
  journal={Journal of Content, Community \& Communication Amity School of Communication},
  volume={6},
  year={2017}
}

@article{au2022role,
  title={The role of online misinformation and fake news in ideological polarization: barriers, catalysts, and implications},
  author={Au, Cheuk Hang and Ho, Kevin KW and Chiu, Dickson KW},
  journal={Information systems frontiers},
  pages={1--24},
  year={2022},
  publisher={Springer}
}

@article{moravec2019fake,
  title={Fake News on Social Media},
  author={Moravec, Patricia L and Minas, Randall K and Dennis, Alan R},
  journal={MIS quarterly},
  volume={43},
  number={4},
  pages={1343--A13},
  year={2019},
  publisher={JSTOR}
}

@inproceedings{said2017cost,
  title={A cost-effective, fast, and robust annotation tool},
  author={Said, Asaad F and Kashyap, Vinay and Choudhury, Namrata and Akhbari, Farshad},
  booktitle={2017 IEEE applied imagery pattern recognition workshop (AIPR)},
  pages={1--6},
  year={2017},
  organization={IEEE}
}

@article{tan2024large,
  title={Large language models for data annotation and synthesis: A survey},
  author={Tan, Zhen and Li, Dawei and Wang, Song and Beigi, Alimohammad and Jiang, Bohan and Bhattacharjee, Amrita and Karami, Mansooreh and Li, Jundong and Cheng, Lu and Liu, Huan},
  journal={arXiv preprint arXiv:2402.13446},
  year={2024}
}

@article{su2022selective,
  title={Selective annotation makes language models better few-shot learners},
  author={Su, Hongjin and Kasai, Jungo and Wu, Chen Henry and Shi, Weijia and Wang, Tianlu and Xin, Jiayi and Zhang, Rui and Ostendorf, Mari and Zettlemoyer, Luke and Smith, Noah A and others},
  journal={arXiv preprint arXiv:2209.01975},
  year={2022}
}

@article{aimeur2023fake,
  title={Fake news, disinformation and misinformation in social media: a review},
  author={A{\"\i}meur, Esma and Amri, Sabrine and Brassard, Gilles},
  journal={Social Network Analysis and Mining},
  volume={13},
  number={1},
  pages={30},
  year={2023},
  publisher={Springer}
}

@article{kruspe2021changes,
  title={Changes in Twitter geolocations: Insights and suggestions for future usage},
  author={Kruspe, Anna and H{\"a}berle, Matthias and Hoffmann, Eike J and Rode-Hasinger, Samyo and Abdulahhad, Karam and Zhu, Xiao Xiang},
  journal={arXiv preprint arXiv:2108.12251},
  year={2021}
}

@article{chen2022election2020,
	title        = {\# Election2020: the first public Twitter dataset on the 2020 US Presidential election},
	author       = {Chen, Emily and Deb, Ashok and Ferrara, Emilio},
	year         = 2022,
	journal      = {Journal of Computational Social Science},
	publisher    = {Springer},
	pages        = {1--18}
}

@article{cinelli2021dynamics,
  title={Dynamics of online hate and misinformation},
  author={Cinelli, Matteo and Pelicon, Andra{\v{z}} and Mozeti{\v{c}}, Igor and Quattrociocchi, Walter and Novak, Petra Kralj and Zollo, Fabiana},
  journal={Scientific reports},
  volume={11},
  number={1},
  pages={22083},
  year={2021},
  publisher={Nature Publishing Group UK London}
}

@article{tripp2025benchmarking,
  title={Benchmarking AI and human text classifications in the context of newspaper frames: A multi-label LLM classification approach},
  author={Tripp, Alexander},
  journal={Research \& Politics},
  volume={12},
  number={2},
  pages={20531680251332353},
  year={2025},
  publisher={SAGE Publications Sage UK: London, England}
}

@inproceedings{zhang2023needle,
  title={A needle in a haystack: An analysis of high-agreement workers on MTurk for summarization},
  author={Zhang, Lining and Mille, Simon and Hou, Yufang and Deutsch, Daniel and Clark, Elizabeth and Liu, Yixin and Mahamood, Saad and Gehrmann, Sebastian and Clinciu, Miruna and Chandu, Khyathi Raghavi and others},
  booktitle={Proceedings of the 61st Annual Meeting of the Association for Computational Linguistics (Volume 1: Long Papers)},
  pages={14944--14982},
  year={2023}
}

@article{kennedy2022repeat,
  title={Repeat spreaders and election delegitimization: A comprehensive dataset of misinformation tweets from the 2020 US election},
  author={Kennedy, Ian and Wack, Morgan and Beers, Andrew and Schafer, Joseph S and Garcia-Camargo, Isabella and Spiro, Emma S and Starbird, Kate},
  journal={Journal of Quantitative Description: Digital Media},
  volume={2},
  year={2022}
}

@article{torabi2019big,
  title={Big Data and quality data for fake news and misinformation detection},
  author={Torabi Asr, Fatemeh and Taboada, Maite},
  journal={Big data \& society},
  volume={6},
  number={1},
  pages={2053951719843310},
  year={2019},
  publisher={SAGE Publications Sage UK: London, England}
}

@inproceedings{pelrine2021surprising,
  title={The surprising performance of simple baselines for misinformation detection},
  author={Pelrine, Kellin and Danovitch, Jacob and Rabbany, Reihaneh},
  booktitle={Proceedings of the web conference 2021},
  pages={3432--3441},
  year={2021}
}

@inproceedings{pinto2025tracking,
  title={Tracking the 2024 US presidential election chatter on Tiktok: a public multimodal dataset},
  author={Pinto, Gabriela and Bickham, Charles and Salkar, Tanishq and Menezes, Joyston and Luceri, Luca and Ferrara, Emilio},
  booktitle={Companion Proceedings of the ACM on Web Conference 2025},
  pages={773--776},
  year={2025}
}

@inproceedings{10.1145/3701716.3715521,
author = {Choi, Eun Cheol and Balasubramanian, Ashwin and Qi, Jinhu and Ferrara, Emilio},
title = {Limited Effectiveness of LLM-based Data Augmentation for COVID-19 Misinformation Stance Detection},
year = {2025},
isbn = {9798400713316},
publisher = {Association for Computing Machinery},
address = {New York, NY, USA},
url = {https://doi.org/10.1145/3701716.3715521},
doi = {10.1145/3701716.3715521},
booktitle = {Companion Proceedings of the ACM on Web Conference 2025},
pages = {934–937},
numpages = {4},
location = {Sydney NSW, Australia},
series = {WWW '25}
}

@inproceedings{10.1145/3485447.3512126,
author = {Papakyriakopoulos, Orestis and Goodman, Ellen},
title = {The Impact of Twitter Labels on Misinformation Spread and User Engagement: Lessons from Trump’s Election Tweets},
year = {2022},
isbn = {9781450390965},
publisher = {Association for Computing Machinery},
address = {New York, NY, USA},
url = {https://doi.org/10.1145/3485447.3512126},
doi = {10.1145/3485447.3512126},
booktitle = {Proceedings of the ACM Web Conference 2022},
pages = {2541–2551},
numpages = {11},
location = {Virtual Event, Lyon, France},
series = {WWW '22}
}

@inproceedings{bedard2018satire,
  title={Satire or fake news: Social media consumers' socio-demographics decide},
  author={Bedard, Michele and Schoenthaler, Chianna},
  booktitle={Companion proceedings of the the web conference 2018},
  pages={613--619},
  year={2018}
}

@article{law2021sensationalist,
  title={Sensationalist social media usage by doctors and dentists during Covid-19},
  author={Law, Richard WM and Kanagasingam, Shalini and Choong, Kartina A},
  journal={Digital Health},
  volume={7},
  pages={20552076211028034},
  year={2021},
  publisher={SAGE Publications Sage UK: London, England}
}

@article{gilardi2023chatgpt,
  title={ChatGPT outperforms crowd workers for text-annotation tasks},
  author={Gilardi, Fabrizio and Alizadeh, Meysam and Kubli, Ma{\"e}l},
  journal={Proceedings of the National Academy of Sciences},
  volume={120},
  number={30},
  pages={e2305016120},
  year={2023},
  publisher={National Academy of Sciences}
}

@inproceedings{wang2024explainable,
  title={Explainable fake news detection with large language model via defense among competing wisdom},
  author={Wang, Bo and Ma, Jing and Lin, Hongzhan and Yang, Zhiwei and Yang, Ruichao and Tian, Yuan and Chang, Yi},
  booktitle={Proceedings of the ACM Web Conference 2024},
  pages={2452--2463},
  year={2024}
}

@inproceedings{gligoric2025can,
  title={Can Unconfident LLM Annotations Be Used for Confident Conclusions?},
  author={Gligori{\'c}, Kristina and Zrnic, Tijana and Lee, Cinoo and Candes, Emmanuel and Jurafsky, Dan},
  booktitle={Proceedings of the 2025 Conference of the Nations of the Americas Chapter of the Association for Computational Linguistics: Human Language Technologies (Volume 1: Long Papers)},
  pages={3514--3533},
  year={2025}
}

@inproceedings{wang2024human,
  title={Human-llm collaborative annotation through effective verification of llm labels},
  author={Wang, Xinru and Kim, Hannah and Rahman, Sajjadur and Mitra, Kushan and Miao, Zhengjie},
  booktitle={Proceedings of the 2024 CHI Conference on Human Factors in Computing Systems},
  pages={1--21},
  year={2024}
}

@article{warrens2011cohen,
  title={Cohen’s kappa is a weighted average},
  author={Warrens, Matthijs J},
  journal={Statistical Methodology},
  volume={8},
  number={6},
  pages={473--484},
  year={2011},
  publisher={Elsevier}
}

@book{krippendorff2018content,
  title={Content analysis: An introduction to its methodology},
  author={Krippendorff, Klaus},
  year={2018},
  publisher={Sage publications}
}

@article{10.1145/3359174,
author = {McDonald, Nora and Schoenebeck, Sarita and Forte, Andrea},
title = {Reliability and Inter-rater Reliability in Qualitative Research: Norms and Guidelines for CSCW and HCI Practice},
year = {2019},
issue_date = {November 2019},
publisher = {Association for Computing Machinery},
address = {New York, NY, USA},
volume = {3},
number = {CSCW},
url = {https://doi.org/10.1145/3359174},
doi = {10.1145/3359174},
journal = {Proc. ACM Hum.-Comput. Interact.},
month = nov,
articleno = {72},
numpages = {23}
}

@article{mchugh2012interrater,
  title={Interrater reliability: the kappa statistic},
  author={McHugh, Mary L},
  journal={Biochemia medica},
  volume={22},
  number={3},
  pages={276--282},
  year={2012},
  publisher={Hrvatsko dru{\v{s}}tvo za medicinsku biokemiju i laboratorijsku medicinu}
}

@article{welinder2010multidimensional,
  title={The multidimensional wisdom of crowds},
  author={Welinder, Peter and Branson, Steve and Perona, Pietro and Belongie, Serge},
  journal={Advances in neural information processing systems},
  volume={23},
  year={2010}
}

@article{ZHANG2017254,
title = {Consensus algorithms for biased labeling in crowdsourcing},
journal = {Information Sciences},
volume = {382-383},
pages = {254-273},
year = {2017},
issn = {0020-0255},
doi = {https://doi.org/10.1016/j.ins.2016.12.026},
url = {https://www.sciencedirect.com/science/article/pii/S0020025516320515},
author = {Jing Zhang and Victor S. Sheng and Qianmu Li and Jian Wu and Xindong Wu}
}

@article{wang2015crowdsourcing,
  title={Crowdsourcing label quality: a theoretical analysis},
  author={Wang, Wei and Zhou, Zhi-Hua},
  journal={Science China Information Sciences},
  volume={58},
  number={11},
  pages={1--12},
  year={2015},
  publisher={Springer}
}

@article{song2020validations,
  title={In validations we trust? The impact of imperfect human annotations as a gold standard on the quality of validation of automated content analysis},
  author={Song, Hyunjin and Tolochko, Petro and Eberl, Jakob-Moritz and Eisele, Olga and Greussing, Esther and Heidenreich, Tobias and Lind, Fabienne and Galyga, Sebastian and Boomgaarden, Hajo G},
  journal={Political Communication},
  volume={37},
  number={4},
  pages={550--572},
  year={2020},
  publisher={Taylor \& Francis}
}

@inproceedings{abacha2025medec,
  title={Medec: A benchmark for medical error detection and correction in clinical notes},
  author={Abacha, Asma Ben and Yim, Wen-wai and Fu, Yujuan and Sun, Zhaoyi and Yetisgen-Yildiz, Meliha and Xia, Fei and Lin, Thomas},
  booktitle={Findings of the Association for Computational Linguistics: ACL 2025},
  pages={22539--22550},
  year={2025}
}

@inproceedings{wang2025leveraging,
  title={Leveraging Large Language Models for Review Classification and Rating Estimation of Mental Health Applications},
  author={Wang, Qile and Erqsous, Moath and Khatiwada, Prerana and Karwankar, Abhishek and Alhassan, Fatimah Mohammad and Chandrasekaran, Aishwarya and Abraham, Benita and Lovell, Faith and Ngo, Andrew Anh and Mauriello, Matthew Louis},
  booktitle={Proceedings of the International AAAI Conference on Web and Social Media},
  volume={19},
  pages={2017--2029},
  year={2025}
}

@article{wang2025lata,
  title={LATA: A Pilot Study on LLM-Assisted Thematic Analysis of Online Social Network Data Generation Experiences},
  author={Wang, Qile and Erqsous, Moath and Barner, Kenneth E and Mauriello, Matthew Louis},
  journal={Proceedings of the ACM on Human-Computer Interaction},
  volume={9},
  number={2},
  pages={1--28},
  year={2025},
  publisher={ACM New York, NY, USA}
}

\appendix

\section{Exploratory Data Analysis}
\label{EDA}

\begin{figure}[!bh]
    \centering
    \includegraphics[width=0.9\columnwidth]{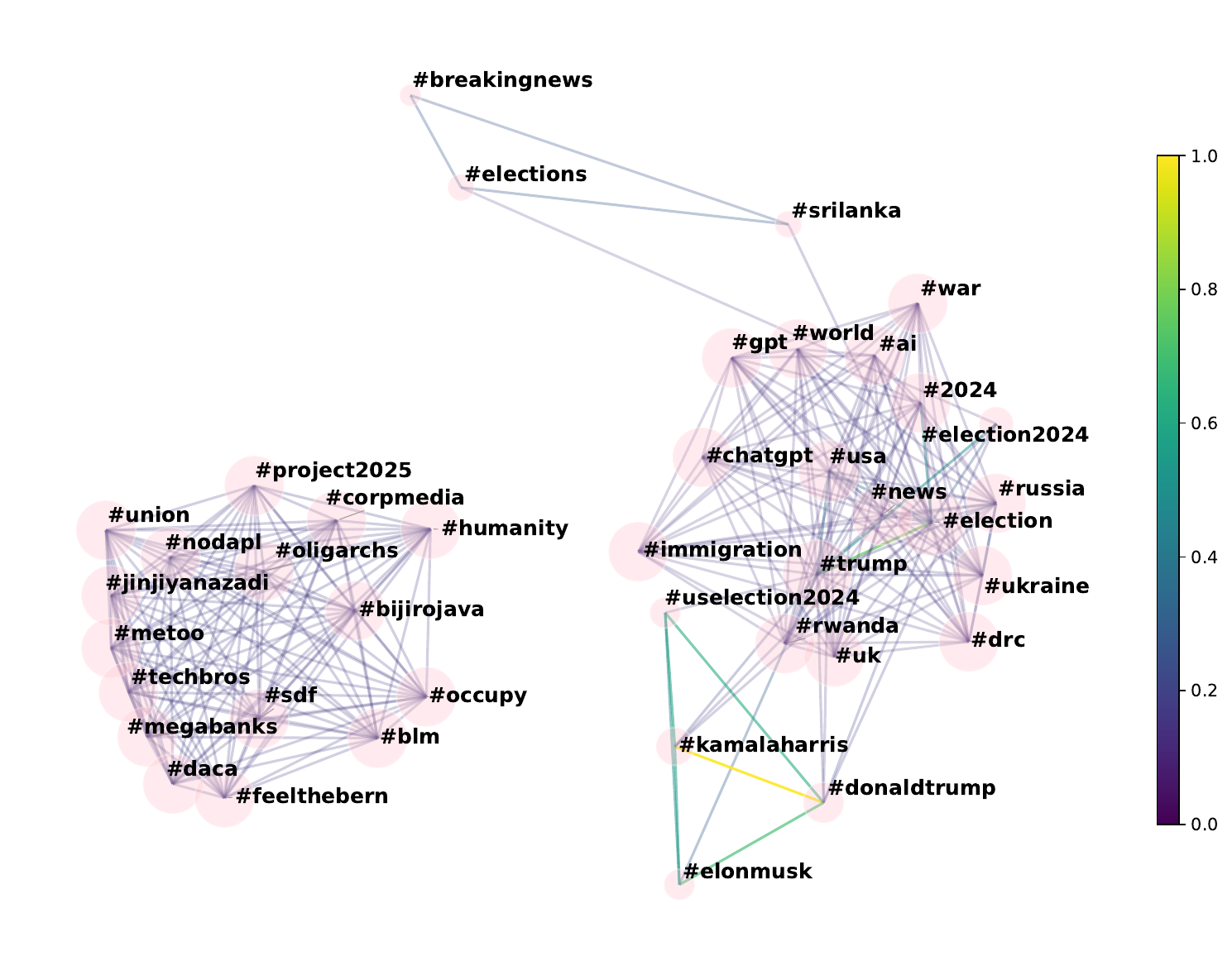}
    \caption{Top 40 hashtag pairs from the 500 most frequent co-occurrences. Node size reflects hashtag frequency; edge color indicates co-occurrence strength.}
\label{fig:occurence}
\end{figure}

\textbf{Co-occurence. }To better understand the thematic landscape of this dataset, we constructed a hashtag co-occurrence graph (Figure~\ref{fig:occurence}).
The visualization effectively shows two prominent, yet distinct, communities of highly correlated hashtags, reflecting divergent thematic discussions on social media. The left cluster predominantly encompasses hashtags associated with progressive activism and social justice, such as \texttt{\#project2025}, \texttt{\#metoo}, and \texttt{\#blm}. Conversely, the right cluster is characterized by a strong focus on contemporary political discourse, technological advancements, and international affairs, highlighted by terms like \texttt{\#ukraine}, \texttt{\#ai}, \texttt{\#immigration}, and political figures such as \texttt{\#trump} and \texttt{\#elonmusk}. The color-coded edges provide insight into the correlation strength, with brighter lines indicating stronger connections, thereby illustrating the varying degrees of thematic cohesion within and between these identified communities.

\begin{figure}[!htbp]
    \centering
    \includegraphics[width=0.9\columnwidth]{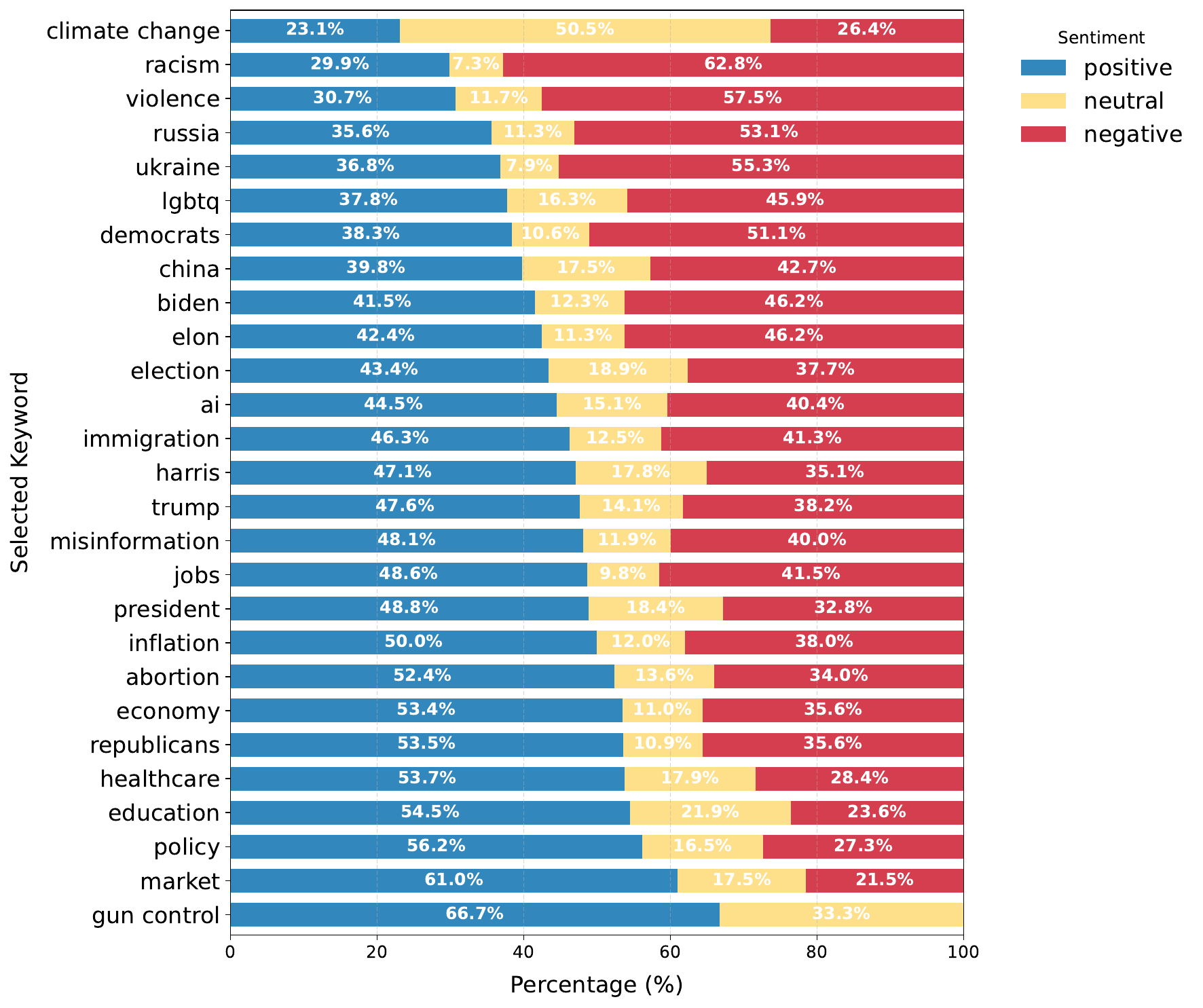}
    \caption{Sentiment distribution of posts containing selected keywords.}
\label{fig:sentimentkey}
\end{figure}

\textbf{Term Polarity.}
Using the sentiment score generated from VADER,~\footnote{https://pypi.org/project/vaderSentiment/} Figure \ref{fig:sentimentkey} illustrates the sentiment distribution of posts related to various selected keywords. To remove sentiment bias from keywords like ``violence,'' we recompute the sentiment score after excluding them from the original text. Each bar reflects the proportion of positive, neutral, and negative sentiments for a given term.  ``gun control" shows the highest positive sentiment at 66.7\%, followed by ``market" (62.6\%) and ``policy" (56.3\%). On the other hand, ``racism" has the highest negative sentiment (62.4\%), alongside ``violence" (57.5\%) and ``ukraine" (56.4\%). Keywords like ``climate change" and ``healthcare" display more balanced sentiment distributions. Terms such as ``trump," ``biden," and ``republicans" show mixed sentiment.
The average sentiment scores also vary across U.S. states, with New Mexico, Wyoming, and New Hampshire displaying the highest positive sentiment, while West Virginia, Kansas, and Oklahoma show the lowest.

\begin{figure}[!htpb]
    \centering
\includegraphics[width=0.99\columnwidth]{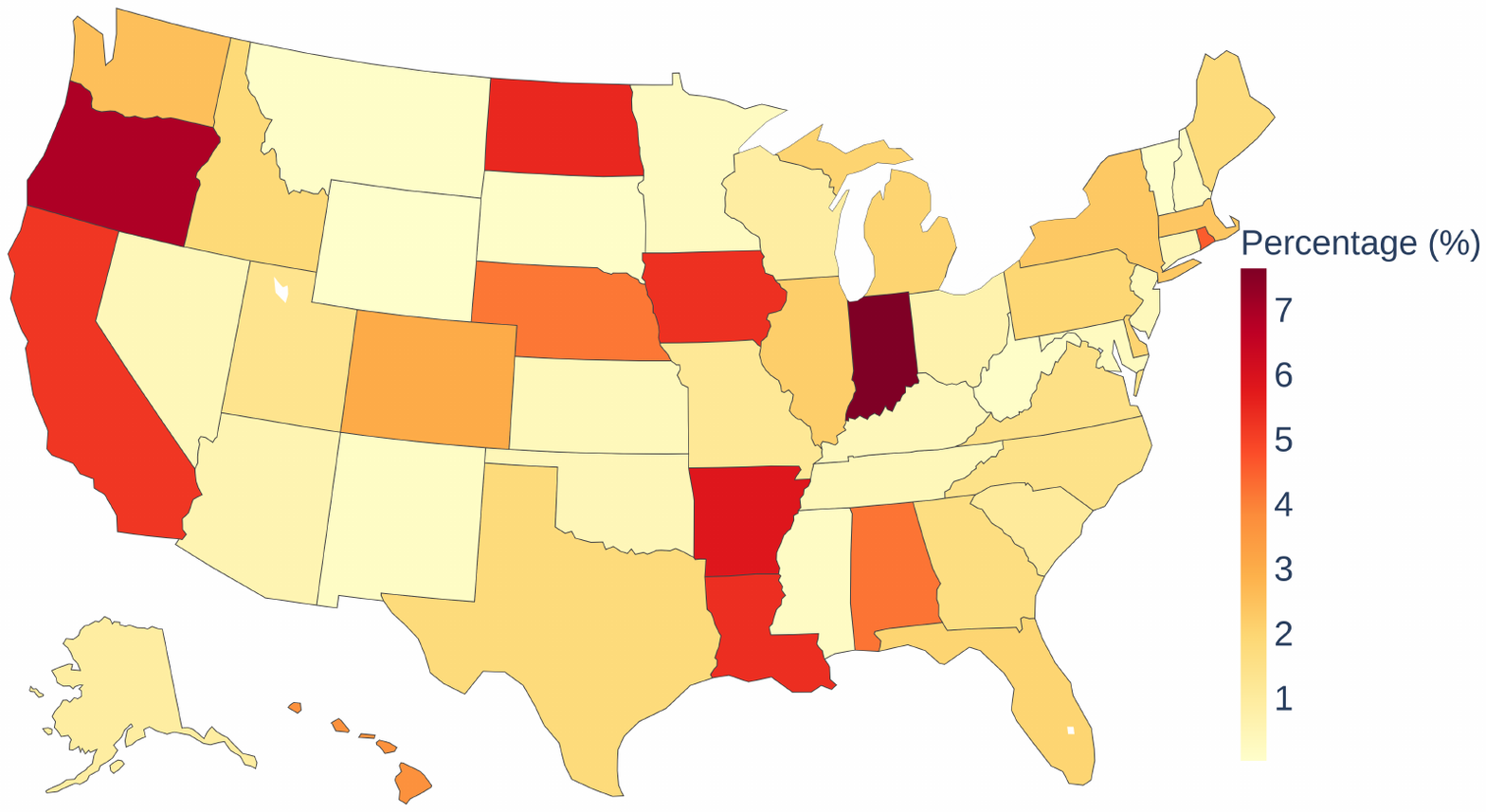}
    \caption{Distribution of Posts Across U.S. States}
\label{fig:posts}
\end{figure}

\textbf{Location.} We also analyzed the post distribution across states to understand regional patterns and trends. Figure \ref{fig:posts} shows Indiana has the highest representation at 7.67\%, followed by Oregon (6.99\%), North Dakota (5.54\%), Louisiana (5.45\%), and California (5.35\%). The data show diverse user participation across states, with relatively high activity in the Midwest and Southern states.

\begin{figure}[!htbp]
\centering
    \centering
  \includegraphics[width=\linewidth]{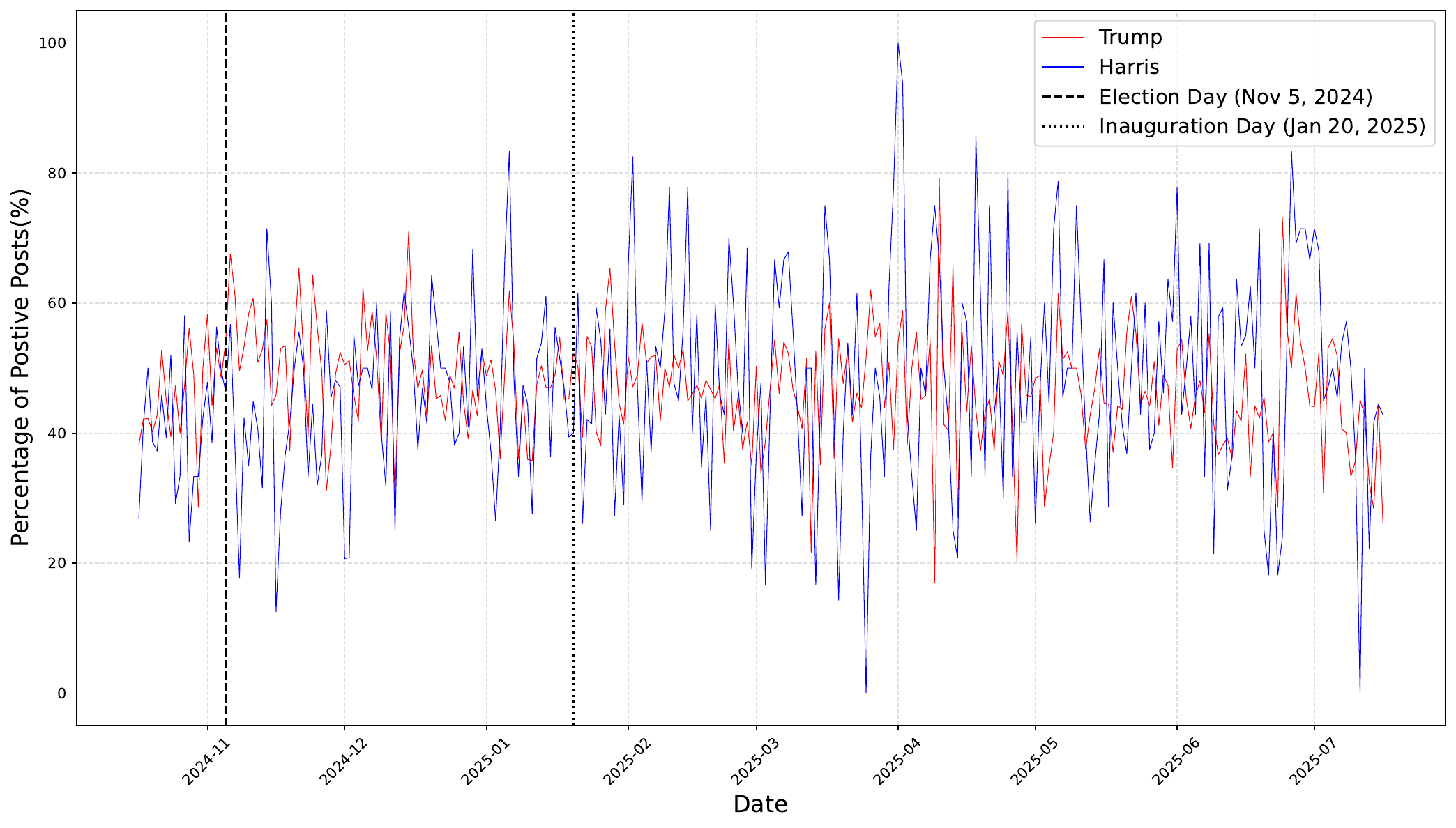} 
    \caption{Temporal Percentage Change in Positive Sentiment for Posts Mentioning ``Trump'' and ``Harris''}
    \label{fig:sentiment}    
\end{figure}

\textbf{Temporal Sentiment.} We were also interested in examining how positive sentiment toward ``Trump'' and ``Harris'' changed over time. Figure \ref{fig:sentiment} illustrates the percentage change, either increasing or decreasing, relative to the total discussion about each candidate on a daily basis. 
Peaks and fluctuations reflect changes in public perception driven by significant events like the Election and Inauguration Day. Overall, Trump shows larger variations, ranging from a high of over 80\% to a low of less 15\%.

\begin{figure*}[!hbtp]
    \centering
    \includegraphics[width=1\linewidth]{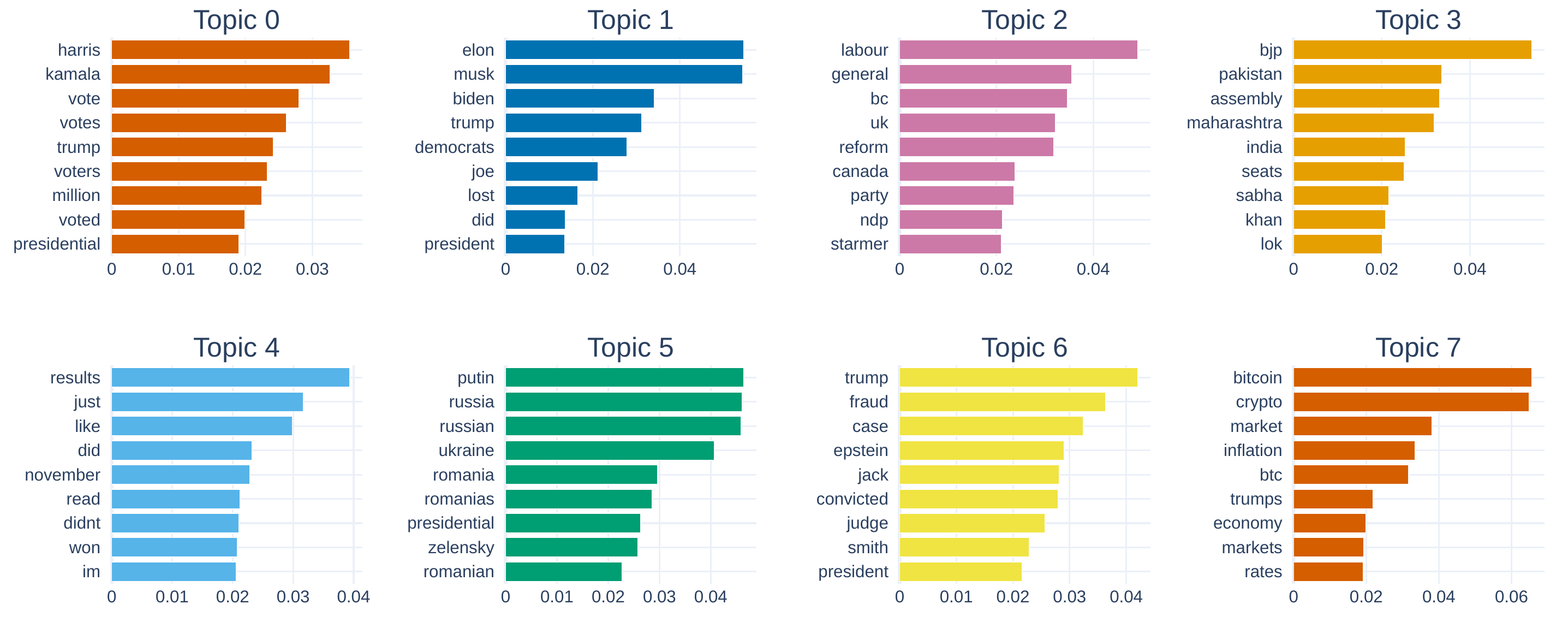}
    \caption{Top 8 discovered topics and their most salient keywords identified.}
    \label{fig:topics}
\end{figure*}

\begin{figure*}[!h]
  \centering
  \begin{subfigure}[t]{0.48\textwidth}
    \centering
    \includegraphics[width=\linewidth]{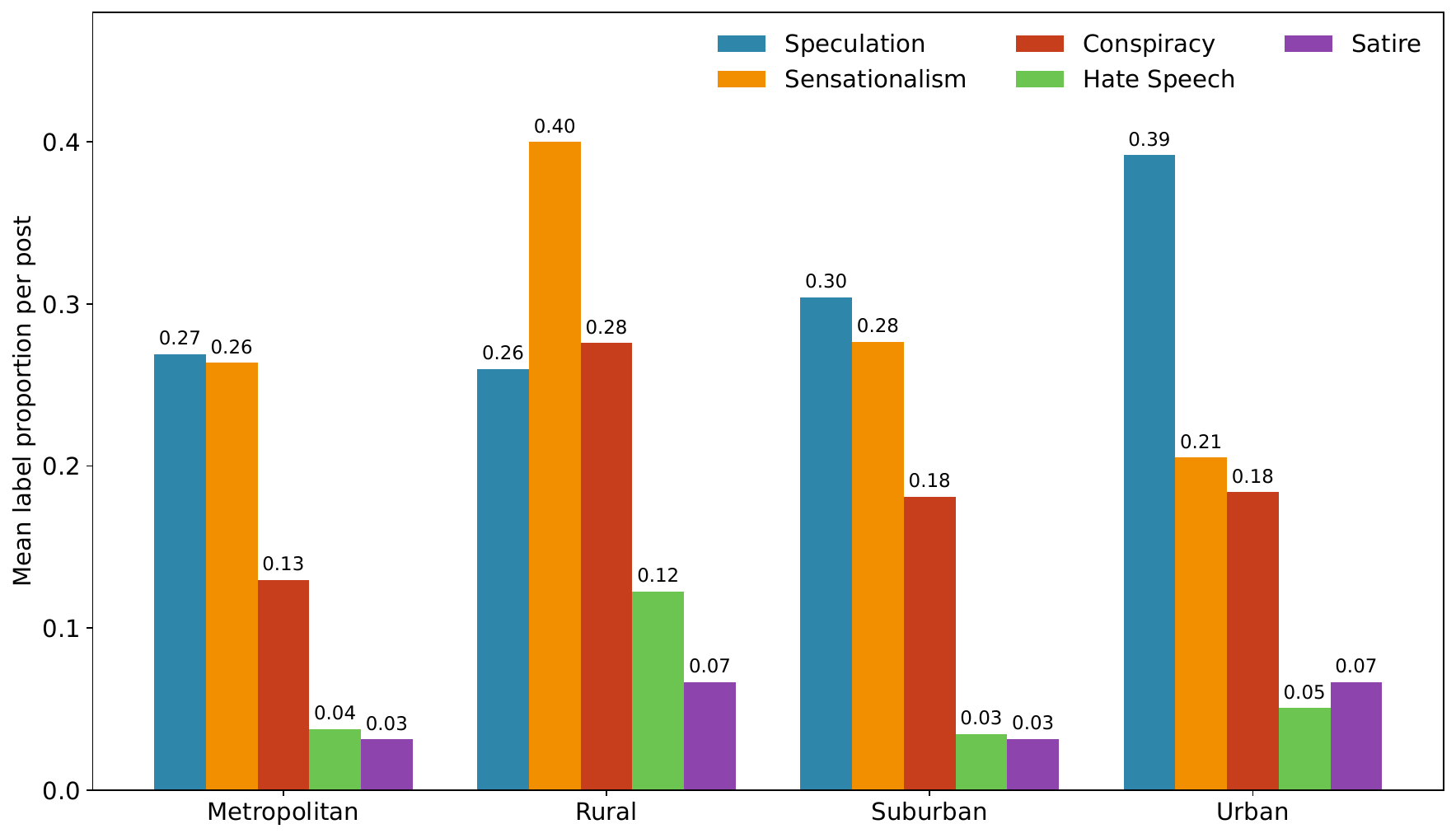}
    \caption{Mean label proportions by \textbf{Area}}
  \end{subfigure}\hfill
  \begin{subfigure}[t]{0.48\textwidth}
    \centering
    \includegraphics[width=\linewidth]{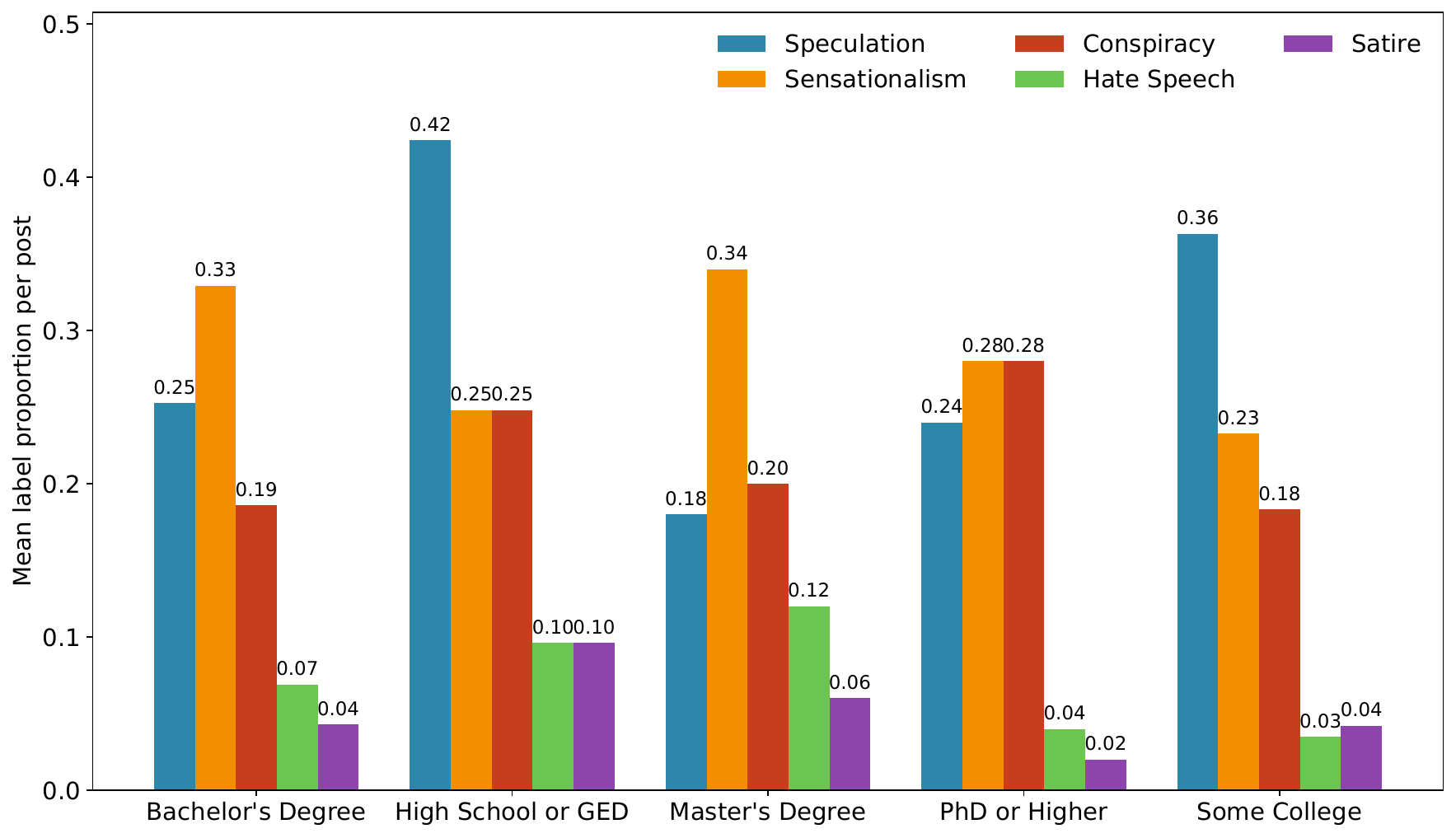}
    \caption{Mean label proportions by \textbf{Education}}
  \end{subfigure}

  \vspace{0.8em}

  \begin{subfigure}[t]{0.48\textwidth}
    \centering
    \includegraphics[width=\linewidth]{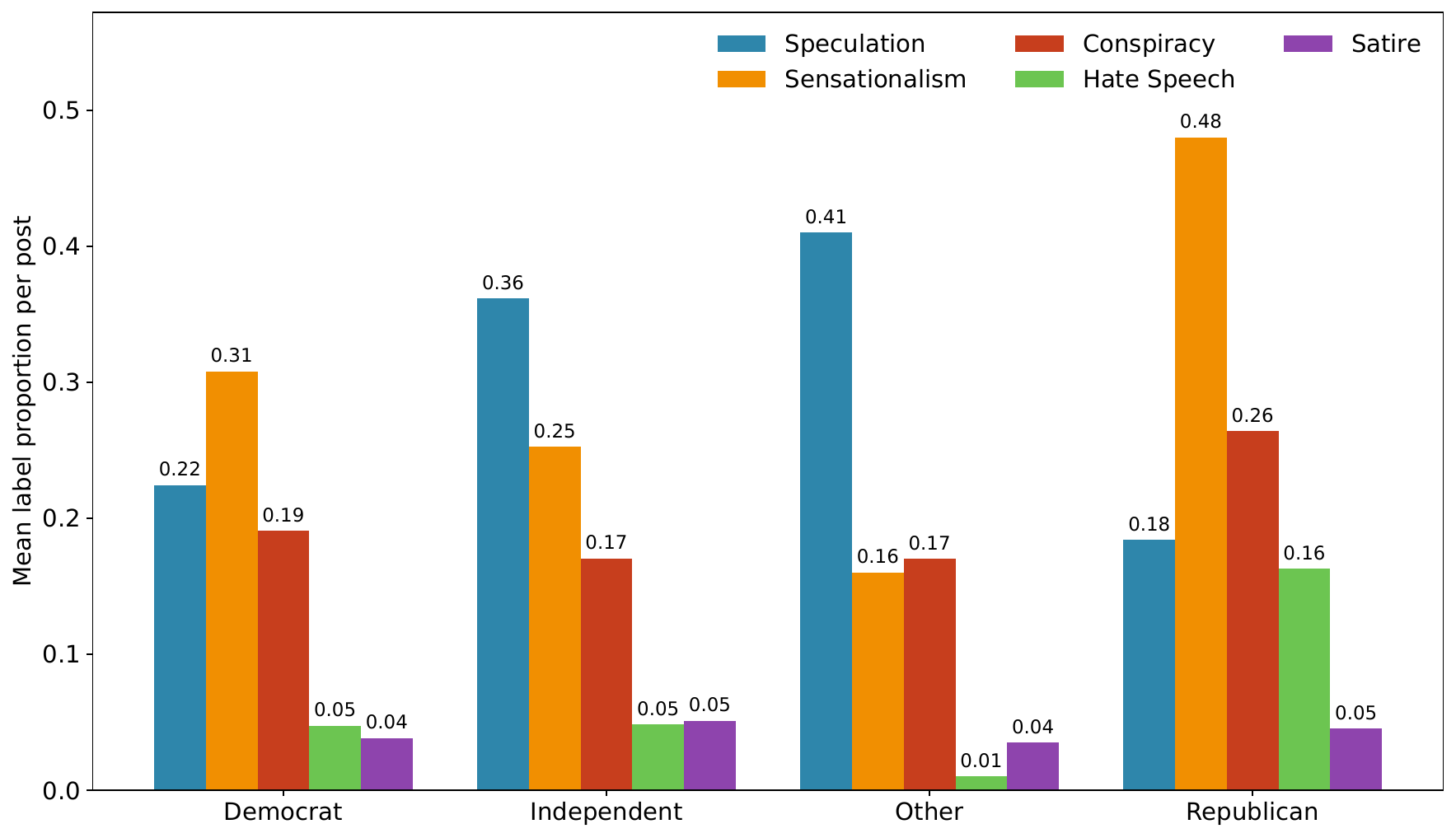}
    \caption{Mean label proportions by \textbf{Political Affiliation}}
  \end{subfigure}\hfill
  \begin{subfigure}[t]{0.48\textwidth}
    \centering
    \includegraphics[width=\linewidth]{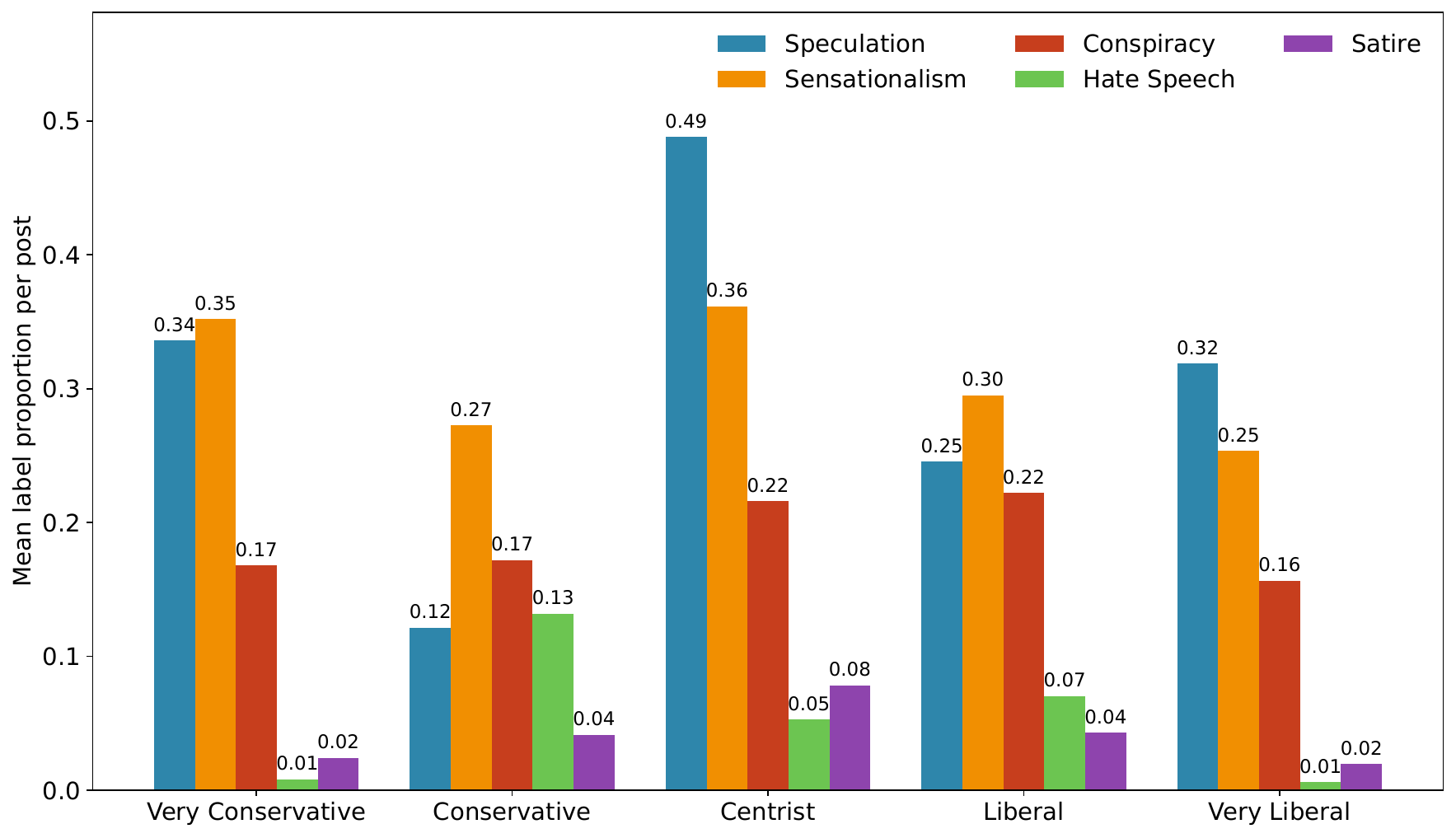}
    \caption{Mean label proportions by \textbf{Ideology}}
  \end{subfigure}

  \caption{\textbf{Mean label proportions across demographic groups.}
  Each bar shows the average proportion of posts within a demographic subgroup that received a given label.
  These descriptive plots illustrate how labeling frequencies vary across (a)~area of residence, (b)~education level,
  (c)~political affiliation, and (d)~self-reported ideology.
  Patterns generally echo the small but significant differences detected by the \ensuremath{\chi^{2}} tests in Section~\ref{Demographic_Variation}.}
  \label{fig:means_by_demographics}
\end{figure*}

\begin{figure*}[!hbtp]
  \centering
  \includegraphics[width=1.10\linewidth]{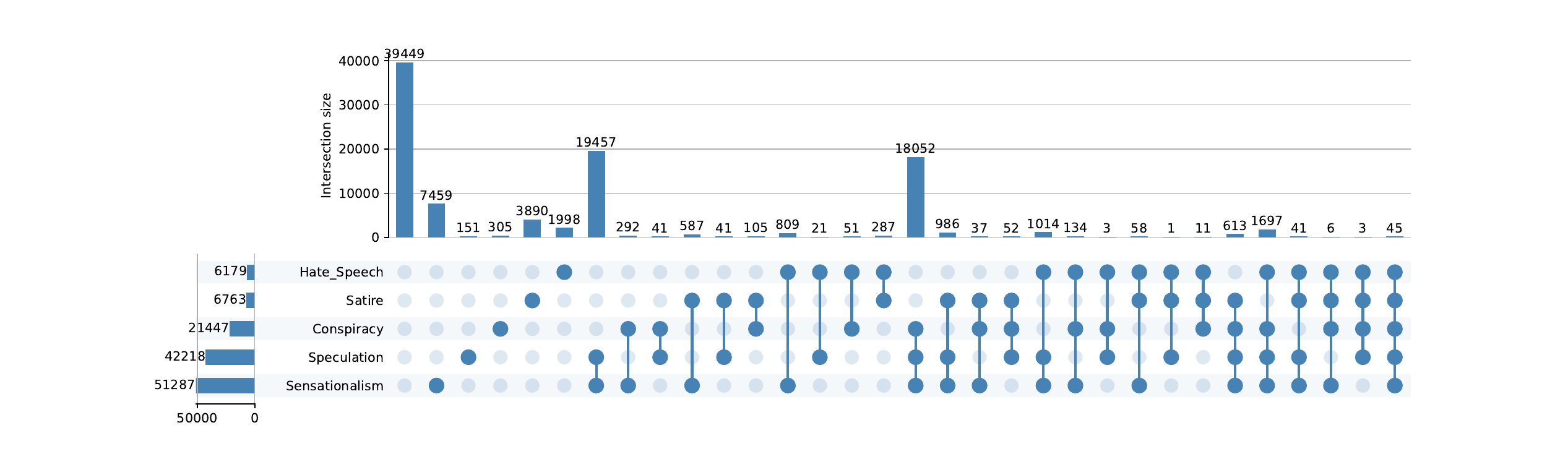}
  \caption{Distribution of multi-label annotations across five harmful content categories with intersection sizes and marginal frequencies of each label.}
  \label{fig:concurrent_labels}
\end{figure*}

\begin{table*}[!tbhp]
\centering
\small
\caption{Pairwise IRR comparison. LLM values are computed from all 15 possible pairs among six LLMs, while human agreement is derived from 193 groups of human annotators.  Values reported in (mean $\pm$ SD, min–max).}
\label{tab:agreement}
\begin{tabular}{lcc|cc}
\toprule
 & \multicolumn{2}{c}{\textbf{Percent Agreement (\%)}} & \multicolumn{2}{c}{\textbf{Cohen's $\kappa$}} \\
\cmidrule(lr){2-3} \cmidrule(lr){4-5}
Category & LLMs & Humans & LLMs & Humans \\
\midrule
Speculation
& 78.82 $\pm$ 5.24 (70.04--85.55)
& 67.89 $\pm$ 9.82 (41.33--86.67)
& 0.58 $\pm$ 0.09 (0.43--0.70)
& 0.21 $\pm$ 0.17 (-0.08--0.57) \\

Sensationalism
& 82.83 $\pm$ 2.72 (77.38--86.78)
& 68.33 $\pm$ 9.76 (49.33--89.33)
& 0.60 $\pm$ 0.07 (0.48--0.71)
& 0.23 $\pm$ 0.14 (0.01--0.65) \\

Conspiracy
& 91.01 $\pm$ 1.76 (87.80--94.70)
& 82.94 $\pm$ 6.42 (68.00--94.67)
& 0.75 $\pm$ 0.05 (0.66--0.84)
& 0.43 $\pm$ 0.18 (0.08--0.75) \\

Hate Speech
& 94.75 $\pm$ 1.30 (92.15--97.33)
& 91.34 $\pm$ 5.58 (81.33--97.33)
& 0.62 $\pm$ 0.08 (0.48--0.73)
& 0.32 $\pm$ 0.18 (-0.02--0.65) \\

Satire
& 94.22 $\pm$ 1.39 (91.23--96.72)
& 92.06 $\pm$ 5.69 (70.67--97.33)
& 0.53 $\pm$ 0.09 (0.35--0.66)
& 0.16 $\pm$ 0.18 (-0.04--0.65) \\
\bottomrule
\end{tabular}
\end{table*}

\textbf{Topic Modeling} We applied neural topic modeling using \textit{all-mpnet-base-v2} embeddings \footnote{\url{https://huggingface.co/sentence-transformers/all-mpnet-base-v2}} after removing the search keywords ``2024'' and ``election'' to avoid trivial topical separation. To select an optimal parameter, we conducted a sensitivity analysis of the minimum topic size parameter and found that 100 provides a stable solution with 35 topics and a lower outlier rate, balancing interpretability and coverage while avoiding topic fragmentation or excessive merging. We then applied post-hoc topic reduction to fix the final number of topics at 30, excluding the outlier category. Figure \ref{fig:topics} presents the top eight topics, covering discourse on U.S. presidential candidates (Topic 0), UK politics and electoral reform (Topic 2), South Asian and regional politics (Topic 3), the Russia–Ukraine conflict and leadership (Topic 5), as well as cryptocurrency and financial markets (Topic 7).

\section{Influence of Demographic Factors}

Figure~\ref{fig:means_by_demographics}
shows how posts were marked across demographic groups. 
Annotators with a high school or GED marked more posts as Speculation and Sensationalism, whereas those with advanced degrees labeled fewer posts overall. Posts marked as Speculation peaked among Independents and Centrists, while conservatives more often marked posts as Hate Speech. Across all subgroups, posts marked as Satire remained consistently rare, highlighting its limited use and potentially ambiguous boundaries compared to other categories. These patterns show that demographic background subtly influences how individuals perceive and categorize online harmful content.

\begin{figure}[!hbtp]
  \centering
  \includegraphics[width=1.0\linewidth]{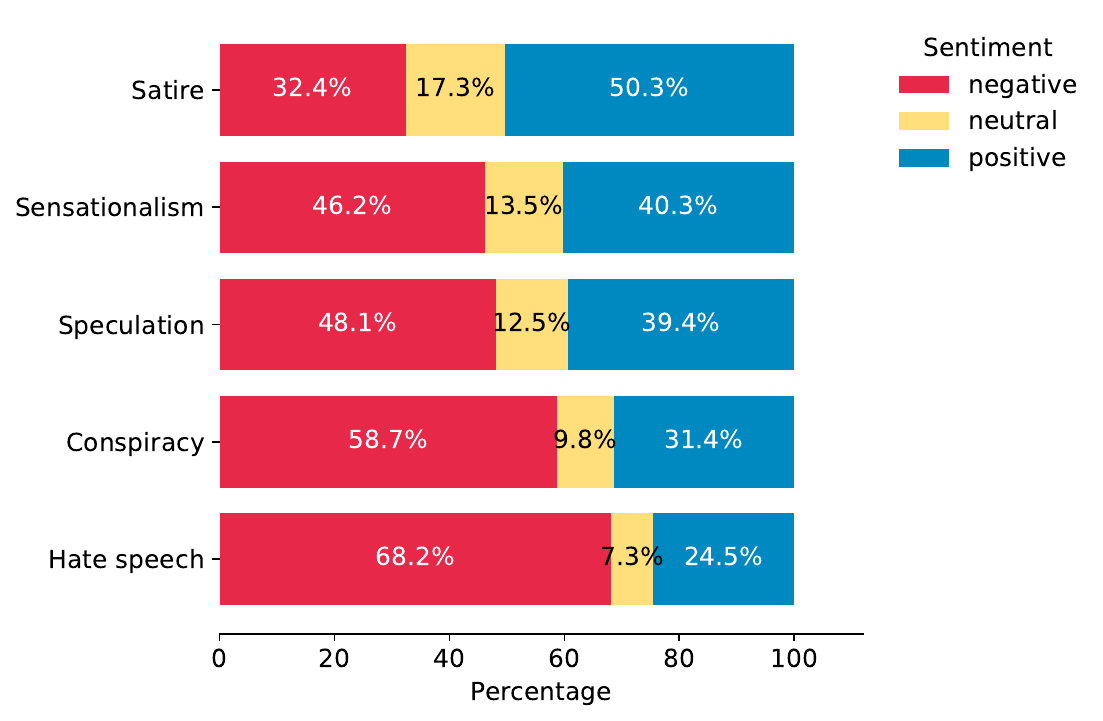}
  \caption{Sentiment distribution across harmful content categories.}
  \label{fig:sentiment_per_annotation}
\end{figure}

\begin{figure*}[!ht]
    \centering
    \begin{subfigure}{0.49\linewidth}
        \centering
        \includegraphics[width=0.85\linewidth]{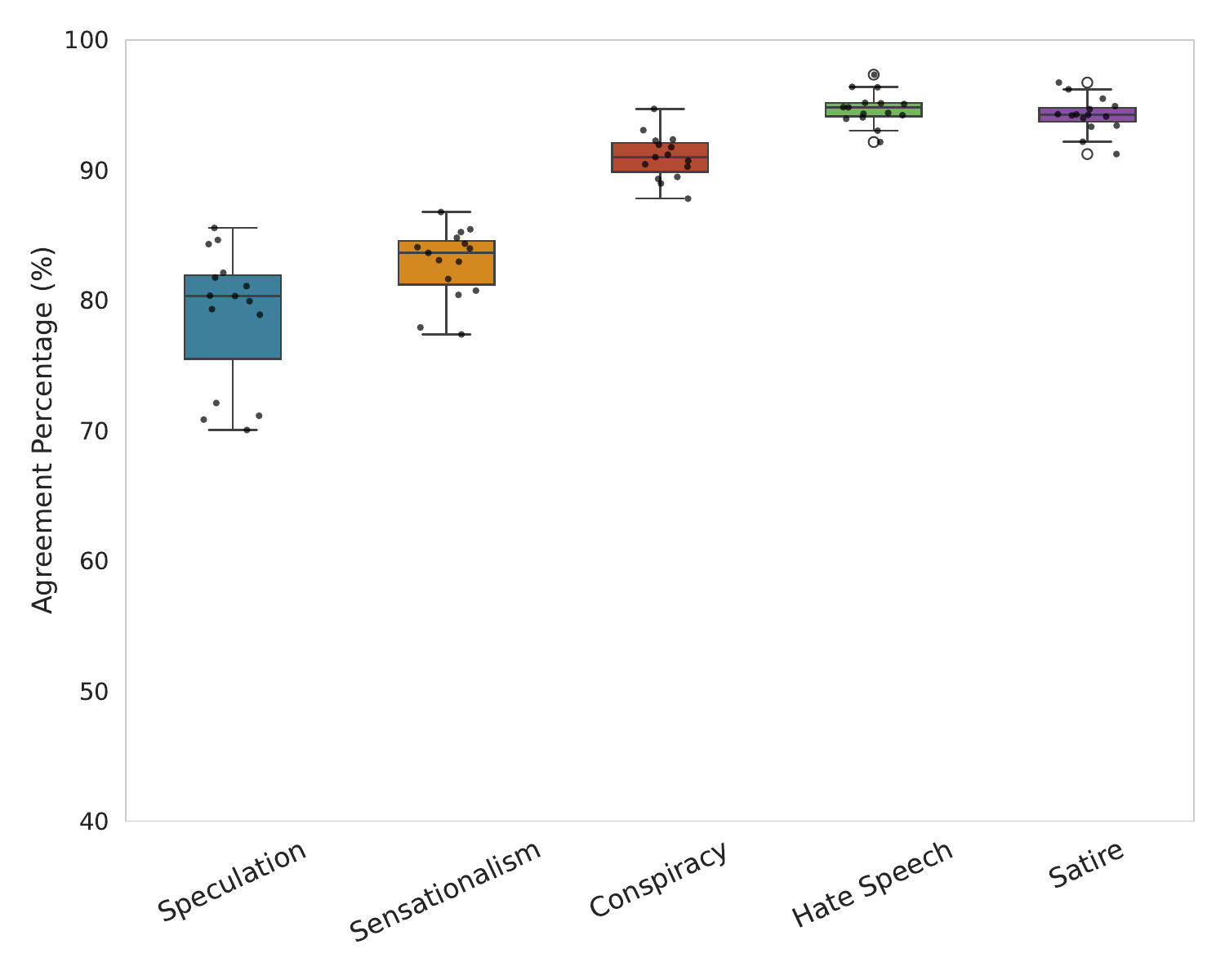} 
         \caption{\centering LLM (n = 15)}
        \label{fig:IRR_LLM_percentage_agreement}
    \end{subfigure}
    \hfill
    \begin{subfigure}{0.49\linewidth}
        \centering
        \includegraphics[width=0.85\linewidth]{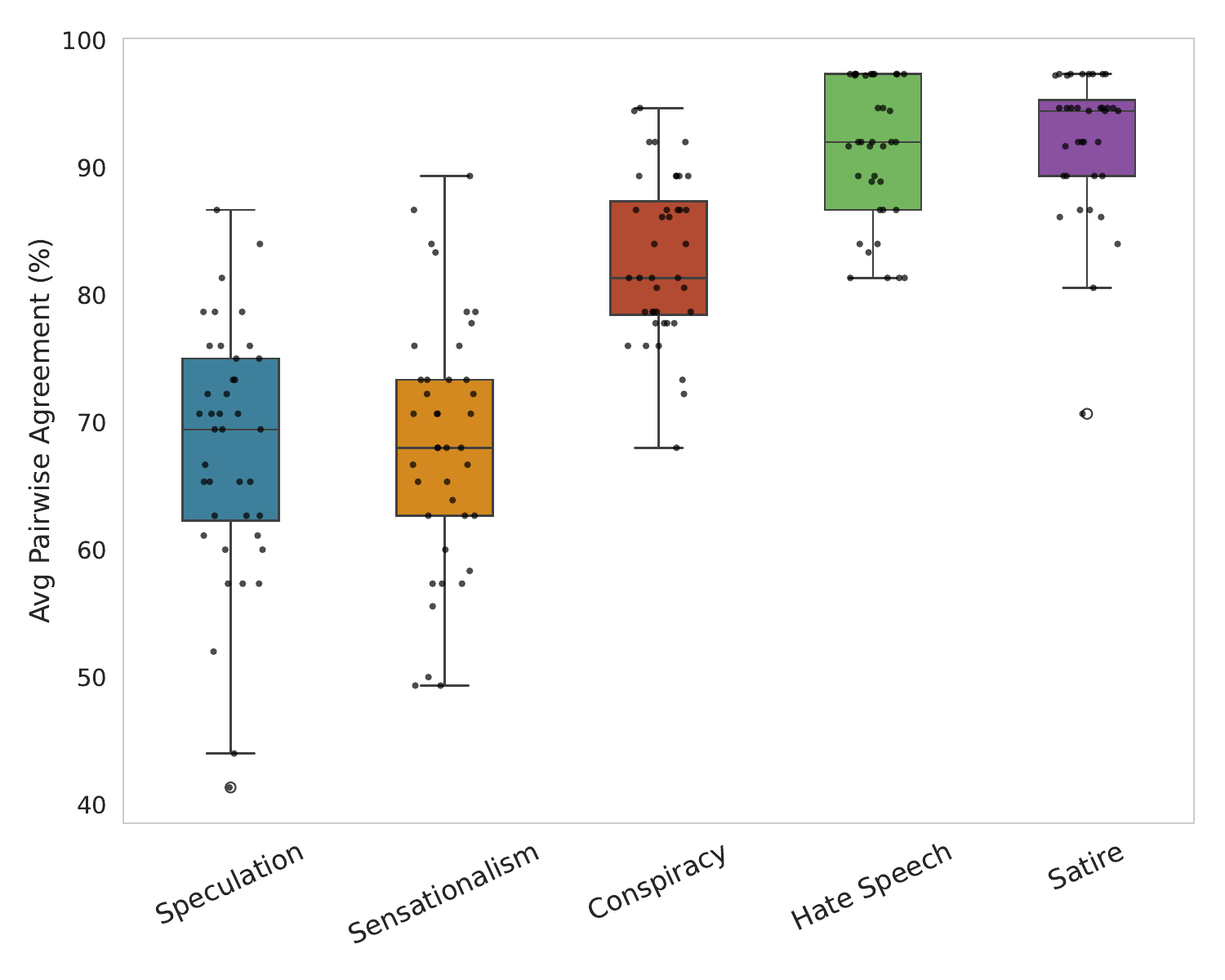}
        \caption{\centering MTurk Workerrs (n = 193)}
        \label{fig:IRR_AMT_percentage_agreement}
    \end{subfigure}
    \caption{\centering Percentage Agreement (\%) IRR Comparison.}
    \label{fig:IRR_Percentage_combined}
\end{figure*}

\begin{figure*}[!ht]
    \centering
    \begin{subfigure}{0.49\linewidth}
        \centering
        \includegraphics[width=0.85\linewidth]{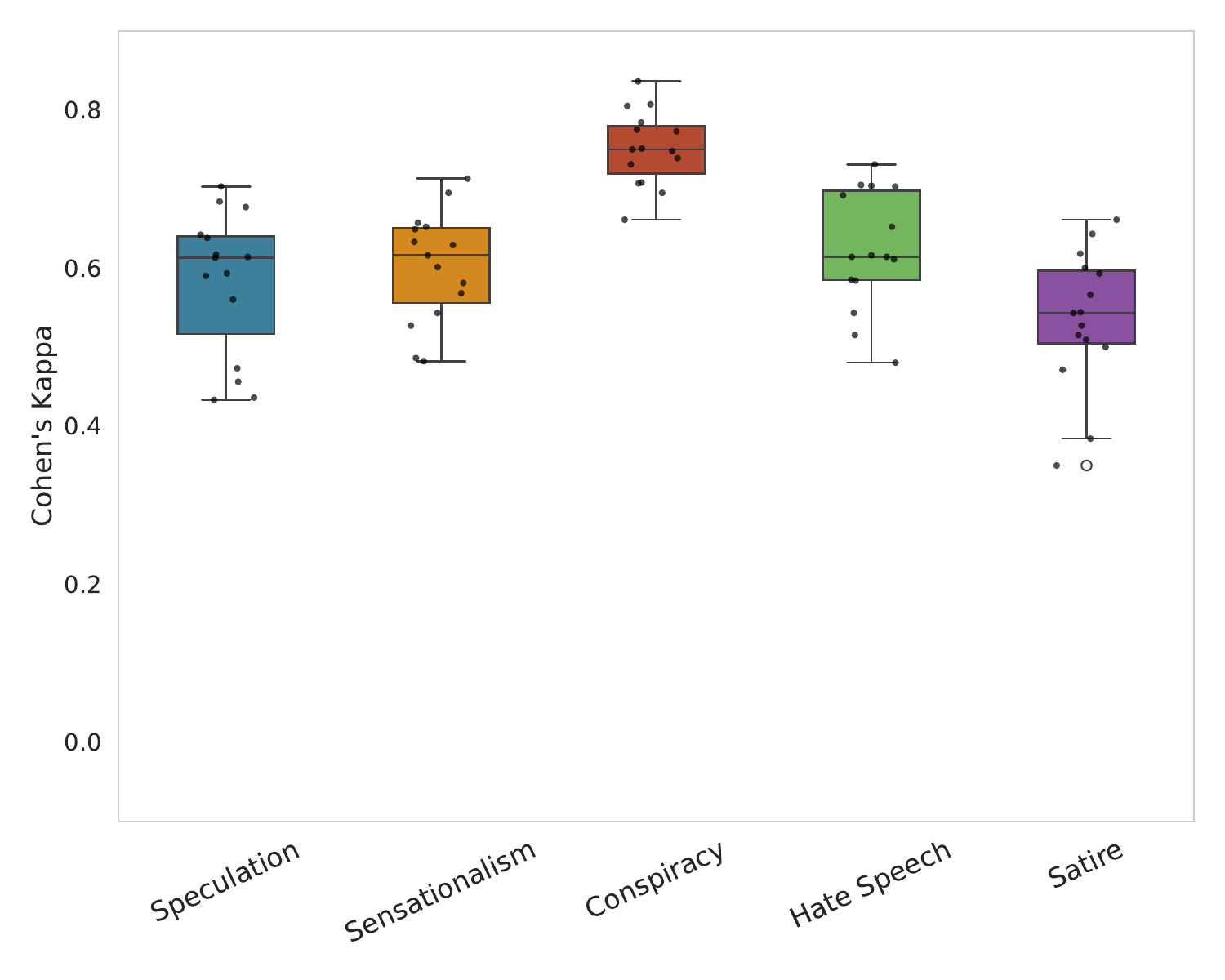} 
        \caption{\centering LLM  (n=15)}
        \label{fig:IRR_LLM_cohen_kappa}
    \end{subfigure}
    \hfill
    \begin{subfigure}{0.49\linewidth}
        \centering
        \includegraphics[width=0.85\linewidth]{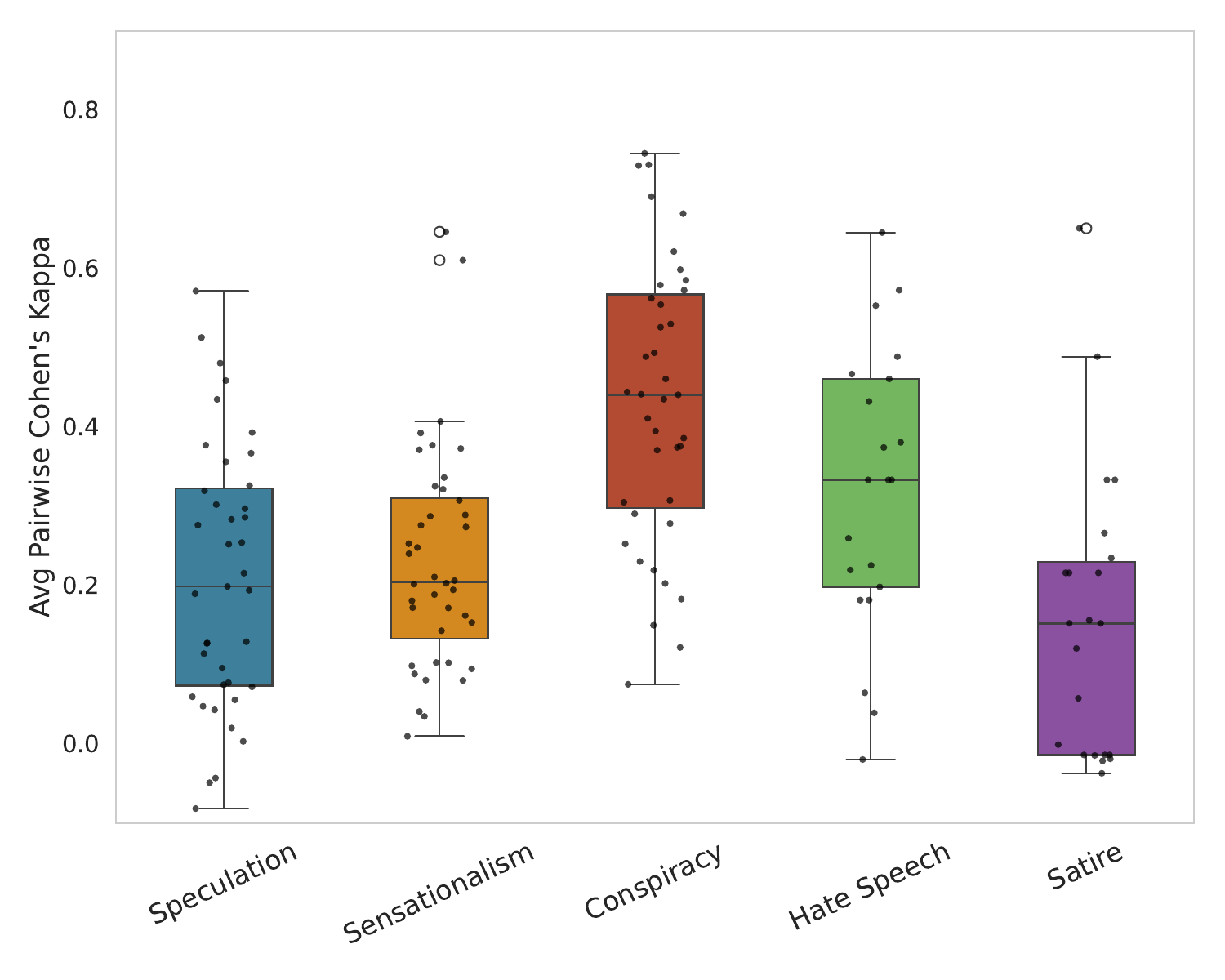}
        \caption{\centering MTurk Workers (n = 193)}
        \label{fig:IRR_AMT_cohen_kappa}
    \end{subfigure}
    \caption{\centering Cohen's Kappa ($\kappa$) IRR Comparison.}
    \label{fig:IRR_cohen_kappa_combined}
\end{figure*}

\section{Pairwise IRR Comparison}

We include additional internal IRR comparisons in Figure \ref{fig:IRR_Percentage_combined}, which shows the percentage of agreement, and Figure \ref{fig:IRR_cohen_kappa_combined}, which presents pairwise Cohen’s Kappa. Detailed measurements are reported in Table \ref{tab:agreement}. Both pairwise IRR measurements exhibit similar trends and patterns for LLMs and MTurk workers across all categories. LLMs demonstrate more consistent agreement compared to human annotators.

\section{Multi Label Distribution}
We examine how harmful content labels co-occur across posts and report the results in Figure \ref{fig:concurrent_labels}. In total, 59.62\% of posts receive at least one label, and 22.20\% receive at least two labels. Sensationalism frequently co-occurs with Speculation and, in many cases, with Conspiracy in addition. This pattern highlights the prevalence of single-category harms and underscores the need to support multi-label analysis for complex content.

Additionally, we present the sentiment distribution for each annotation category in Figure \ref{fig:sentiment_per_annotation}. Overall, more severe categories such as Hate speech and Conspiracy show the highest proportion of negative sentiment, suggesting that posts labeled under these categories tend to convey stronger negative emotional tone. In contrast, Satire exhibits the highest share of positive sentiment, indicating a more mixed or less hostile narrative context.

Neutral sentiment remains relatively low across all categories, which suggests that discussions around these topics are often emotionally polarized rather than purely informational. The gradual shift from highly negative sentiment in Hate speech to more positive sentiment in Satire highlights how different narrative frames may shape emotional expression and engagement within online political discourse.

\end{document}